\documentclass[fleqn,usenatbib]{mnras}
\usepackage{newtxtext,newtxmath}

\usepackage[utf8]{inputenc}
\usepackage[T1]{fontenc}
\usepackage{ae,aecompl}
\usepackage{graphicx}	
\usepackage{amsmath}	
\usepackage{amssymb}	
\usepackage[normalem]{ulem}
\usepackage{multicol}
\usepackage{pdflscape}
\usepackage{lineno}

\newcommand{\noop}[1]{}

\usepackage{longtable}
\usepackage{array}
\usepackage{rotating}
\newcommand{\PreserveBackslash}[1]{\let\temp=\\#1\let\\=\temp}
\newcolumntype{C}[1]{>{\PreserveBackslash\centering}p{#1}}
\newcolumntype{R}[1]{>{\PreserveBackslash\raggedleft}p{#1}}
\newcolumntype{L}[1]{>{\PreserveBackslash\raggedright}p{#1}}

\newcommand{\mmode}[1]{\ifmmode{#1}\else{$#1$}\fi}

\newcommand{\Teff}[0]{\mmode{T_\text{eff}}}

\newcommand{\Msolar}[0]{\mmode{\text{M}_{\odot}}}

\newcommand{\Dnu}[0]{\mmode{\Delta\nu}}

\newcommand{\numax}[0]{\mmode{\nu_\text{max}}}

\usepackage{CJKutf8}
\newcommand{\CNnames}[1]{{\begin{CJK}{UTF8}{gbsn}~(#1)~\end{CJK}}}

\newcommand{\Nofstars}{48}
\newcommand{\totalNofstars}{114}
\newcommand{\yucat}{Y18}

\title[High-mass Kepler RGs]{The highest mass Kepler red giants--- I. Global asteroseismic parameters of \Nofstars{} stars}
\author[C. L. Crawford et al.]{
Courtney L. Crawford$^{1}$\thanks{Email: courtney.crawford@sydney.edu.au},
Timothy R. Bedding$^{1}$,
{Yaguang~Li\CNnames{李亚光}}$^{1,2}$,
\newauthor \ Dennis Stello$^{1,3}$,
Daniel Huber$^{1,2}$,
{Jie Yu\CNnames{余杰}}$^{4}$,
K. R. Sreenivas $^{1}$,
\newauthor \ {Tanda Li\CNnames{李坦达}}$^{5,6,7}$,
Emily F. Kerrison$^{1,8,9}$
\\
$^{1}$ Sydney Institute for Astronomy (SIfA), School of Physics, University of Sydney, NSW 2006, Australia\\
$^{2}$ Institute for Astronomy, University of Hawai`i, 2680 Wood-lawn Drive, Honolulu, HI 96822, USA\\
$^{3}$ School of Physics, University of New South Wales, Sydney, NSW 2052, Australia \\
$^{4}$ School of Computing, Australian National University, Acton, ACT 2601, Australia \\
$^{5}$ Institute for Frontiers in Astronomy and Astrophysics, Beijing Normal University, Beijing 102206, China\\
$^{6}$ Department of Astronomy, Beijing Normal University, Beijing 100875, China\\
$^{7}$ School of Physics and Astronomy, The University of Birmingham, UK, B15 2TT\\
$^{8}$ ARC Centre of Excellence for All Sky Astrophysics in 3 Dimensions (ASTRO 3D) \\
$^{9}$ ATNF, CSIRO Space and Astronomy, PO Box 76, Epping, NSW 1710, Australia
}
\date{Accepted 2024 February 7. Received 2024 January 22; in original form 2023 November 5}
\pubyear{2023}
\begin{document}
\label{firstpage}
\pagerange{\pageref{firstpage}--\pageref{lastpage}}
\maketitle

\begin{abstract}
When low- and intermediate-mass stars evolve off the main sequence, they expand and cool into the red giant stages of evolution, which include those associated with shell H burning (the red giant branch), core He burning (the red clump), and shell He burning (the asymptotic giant branch). The majority of red giants have masses < 2 \Msolar{}, and red giants more massive than this are often excluded from major studies. Here we present a study of the highest-mass stars (M > 3.0 \Msolar{}) in the Kepler sample of 16,000 red giants. We begin by re-estimating their global seismic properties with new light curves, highlighting the differences between using the SAP and PDCSAP light curves provided by Kepler. We use the re-estimated properties to derive new mass estimates for the stars, ending with a final sample of \Nofstars{} confirmed high-mass stars. We explore their oscillation envelopes, confirming the trends found in recent works such as low mean mode amplitude and wide envelopes. We find, through probabilistic means, that our sample is likely all core He burning stars. We measure their dipole and quadrupole mode visibilities and confirm that the dipole mode visibility tends to decrease with mass. 

\end{abstract}


\section{Introduction}

Red giants are a late-stage evolutionary phase for most low- to intermediate-mass stars. When these stars evolve off the main sequence after core hydrogen exhaustion, they begin shell hydrogen burning and expand to large radii as they move upwards on the red giant branch (RGB). However, once they become hot enough to ignite helium in their cores, either in a flash or not, they shrink down slightly for core-He burning (CHeB) in what we sometimes call the red clump phase. Eventually, they will once again exhaust their cores, this time of helium, and expand. This time they ascend the asymptotic giant branch (AGB), where they will eventually shed their envelopes and move on to become a white dwarf. In each of the three giant phases, collectively referred to as the red giants, many stars will exhibit stellar oscillations, making them ripe for asteroseismic study.

Contrary to the classical pulsators, whose pulsations are driven by opacity bumps in the stellar atmosphere (the $\kappa$-mechanism), red giant oscillations are driven stochastically by near-surface convection. This is the same mechanism that excites oscillations in the Sun and so red giants fall under the umbrella of ``solar-like'' oscillators, which includes many star types that lie on the cool side of the classical instability strip (see \citealt{Bedding+Kjeldsen2003_review, Chaplin+Miglio2013_review, Jackiewicz2021_review} for reviews). In solar-like pulsators, a range of oscillation frequencies are excited. These are typically described using overall parameters. The two main parameters used to describe solar-like oscillations are \numax{}, the frequency of maximum power of the nearly Gaussian-shaped envelope of the excited frequencies, and \Dnu{}, the approximately equal frequency separation between consecutive radial modes. The excited modes in the star each correspond to certain spherical harmonics, and thus can be separated into radial, dipole, and quadrupole modes, or $\ell$=0, $\ell$=1, and $\ell$=2 modes, respectively. While the radial modes are seen most often in stars, the dipole and quadrupole modes are observable in many cases. Observable dipole modes, in particular, convey interesting information about the near-core region of the star due to their mixed character, i.e. the coupling of acoustic (p-mode) waves in the envelope and buoyancy (g-mode) waves near the core of the star \citep[see reviews by][]{Hekker2017_review, Garcia2019_review_solar_like}. 

While red giants are the late stages of both low- to intermediate-mass stars, observed red giants have a relatively narrow mass range of roughly 0.8 \Msolar{} to 5 \Msolar{} (in single star systems). The vast majority of red giants have lower masses, and very few stars have estimated masses greater than $\sim$ 2.5 \Msolar{} \citep{Pinsonneault2018_apokasc2, Yu2018_keplercatalog, Hon2021_tess_qlp, Mackereth2021_tess_cvz}. Most studies of red giants do not consider masses greater than 3 \Msolar{} due to their rarity. 

The highest-mass red giants are important for a few reasons. Firstly, stars greater than $\sim$2 \Msolar{} will not develop degenerate He cores on the RGB, and thus will not undergo explosive ignition of He (the ``helium flash"). When they do ignite their He core, they will settle onto a slightly different region of the Hertzsprung-Russell diagram (HRD) in what is called the ``secondary clump", as opposed to the ``red clump" for the lower-mass red giants \citep{Girardi1999_secondaryclump}. Secondly, many asteroseismic trends are known to be strong functions of mass, especially the qualities of the envelope of excited modes 
\citep{Kjeldsen2011_amplitudescaling,Huber2011_solarvalues,Mosser2012_powerexcess,Stello2013_pspacingtracks,Kallinger2014_granulation,Yu2018_keplercatalog} and the relative visibility of the different angular degrees of oscillation \citep{Mosser2012_powerexcess,Fuller2015_suppression,Stello2016_visNature,Stello2016_visPASA,Cantiello2016_suppression}. 
The high-mass red giants also provide useful anchor points to other sets of stars, such as their progenitor main-sequence stars---which may often be slowly-pulsating B stars---and the white dwarfs. In clusters, clump stars are useful in determining overall mass loss rates on the RGB \citep{Miglio2012_clustermassloss,Handberg2017_masslosscluster,Stello2016_masslossM67,Howell2022_masslossM4} and therefore the highest-mass secondary clump stars may further improve our understanding of RGB mass loss, as has been studied for the most luminous red giants \citep{Yu2021_luminosrg_massloss}. High-mass red giants should also be fairly young stars with high metallicities, and can therefore probe the dependence of mass loss on metallicity. Finally, while the accuracy of the asteroseismic scaling relations for mass and especially radius have been tested using eclipsing binaries and interferometric measurements \citep{Huber2011_testingscalingCHARA,Brogaard2016_testingscaling,Gaulme2016_testingscaling,Zinn2019_testingscalingrelation_gaia,Brogaard2022_testingscaling}, the highest-mass red giants provide an important opportunity to test these scaling relations at their extremes.
It is particularly important to test the \numax{} scaling relation \citep{Brown1991_numaxscaling, Kjeldsen1995_scalingrelations} for stars with higher masses because it has been observed that they tend to have broader oscillation envelopes (\citealt{Mosser2012_powerexcess, Yu2018_keplercatalog, Kim2021_width}; Sreenivas et al. 2024, submitted). 
For all these reasons, we choose to study in detail those stars at the highest mass end of the Kepler red giant sample, with the express goal of modelling these stars in order to test the scaling relations.

In this paper we present a `legacy' sample of the highest mass Kepler red giant stars ($>3\Msolar$) from the catalogue of \citet{Yu2018_keplercatalog}, which are well-suited for further asteroseismic study. In Section~\ref{sec:sample} we define the sample selection and discuss the methods by which we refine their oscillation and spectroscopic parameters. In Section~\ref{sec:demographics} we present the demographics of this sample, with discussion of potential binarity and contamination in Subsection~\ref{subsec:binarity}, their evolutionary phases in Subsection~\ref{subsec:evol_phase}, their oscillation envelopes in Subsection~\ref{subsec:envelopes}, the mode visibilities in Subsection~\ref{subsec:suppression}, and radial mode glitches in Subsection~\ref{subsec:glitches}. Finally, in Section~\ref{sec:conclusions} we present a summary of our findings from the study of these highest mass stars and our plans for future work.


\section{Target Selection}
\label{sec:sample}

Our aim is to provide a `legacy' sample of the highest mass Kepler red giants near the red clump portion of the HRD. We choose our stars from the catalogue of 16,000 Kepler red giants compiled by \citet{Yu2018_keplercatalog}, hereafter referred to as \yucat. We begin by choosing all stars with quoted red clump corrected masses greater than 3.0 \Msolar{} and at least six quarters observed by Kepler (see Section~\ref{subsec:evol_phase} for justification of the clump assumption). This leaves us with a starting selection of \totalNofstars{} stars. We now take a detailed look at each star and their parameters to confirm the mass estimates in \yucat.

We use the Kepler PDCSAP (Presearch Data Conditioning of Simple Aperture Photometry) light curves taken in long cadence mode ($\approx$30-minute integration) for these \totalNofstars{} stars. We pass their stitched light curves through a Gaussian high-pass filter with a width of 1 $\mu$Hz followed by a 4$\sigma$-clipping. The high-pass filtering accounts for the few quarters where the PDC algorithm did not properly remove the instrumental trends (see \citealt{Stumpe2012_PDCKepler} and \citealt{Kinemuchi2012_demystifyingkepler} for description of the PDC algorithm). The width of this filter is well below any oscillation frequencies in our sample ($>$20 $\mu$Hz) and therefore does not affect the oscillations we are studying. We then use the Lomb-Scargle method \citep{Lomb1976_lombscargle,Scargle1982_lombscargle} to convert these light curves to power spectra to re-derive new estimates of \numax{} and \Dnu{}, taking care that our estimations are self-consistent and precise. In the following sections (\ref{subsec:numax}, \ref{subsec:dnu}, and \ref{subsec:teff}), we describe in detail the methods we used to validate both of these parameters, along with the adopted \Teff{}.

\subsection{\numax{} adjustment}
\label{subsec:numax}

\numax{} itself is not a value that can be defined unambiguously. It measures the frequency of maximum power but, considering the stochastic nature of solar-like oscillations, the power of each oscillation mode may change over time as the modes are continually damped and re-excited. There are also several different ways to measure \numax{}. The most commonly used method is to fit both the background and a Gaussian for the power excess, the mean of which gives the \numax{} estimate (see \citealt{Hekker2011_comparison} and references therein). Another method, used by both \yucat{} and this work, is to again perform a background fit and subtraction, and then heavily smooth the remaining power excess. The peak of this smoothed power excess would then be the \numax{} estimate \citep{Kjeldsen2005_alphacenB,Huber2009_sydpipeline}. Generally, the difference between these two methods is very small, but for some stars they can differ considerably. This certainly impacts estimation of the mass. 

We recalculate the \numax{} for every star using the {\sc pySYD} program \citep{Chontos2022_pysyd}, which is a python adaptation of the {\sc SYD} pipeline \citep{Huber2009_sydpipeline} used by \yucat{}. {\sc pySYD}'s default options are designed to fit the power spectra of cool dwarf stars instead of red giants, and we therefore used a few specific non-default options (see the documentation for explanation of these parameters): \texttt{ind\_width} of 1, \texttt{lower\_bg} of 1, \texttt{smooth\_ps} of 0.5, \texttt{lower\_ex} of 1, \texttt{smooth\_width} of 5, and \texttt{binning} of 0.01. The majority of these parameters correspond to different smoothing parameters used for {\sc pySYD}'s different fitting steps, as well as power spectrum boundaries. We highlight in particular that \texttt{lower\_bg} of 1 indicates we are only fitting the background for $\nu$~>~1~$\mu$Hz, which is the cut-off frequency of our high-pass filter. {\sc pySYD} chooses the number of background models dynamically, and chose 2 profiles for each of our stars. Note that granulation background is fit using a Harvey profile \citep{Harvey1985} which is defined as a Lorentzian, however it is common for pipelines to fit super-Lorentzians or other similar functions to the background (see discussion in \citealt{Mathur2011_granulation}). Both this work and \yucat{} fit the background with profiles proportional to $1/(1+x^2+x^4)$ where x is a linear function of frequency.
{\sc pySYD} outputs both a Gaussian and a smoothed estimate of the \numax{}, as discussed above. For our work we followed \yucat{} in adopting the smoothed estimate. {\sc pySYD} calculates uncertainties by performing the fitting process a large number of times on perturbed power spectra and provides the standard deviations of the posterior distributions \citep{Chontos2022_pysyd}. We used an iteration count of 200 for our stars.

\begin{figure}
    \centering
    \includegraphics[width=\columnwidth]{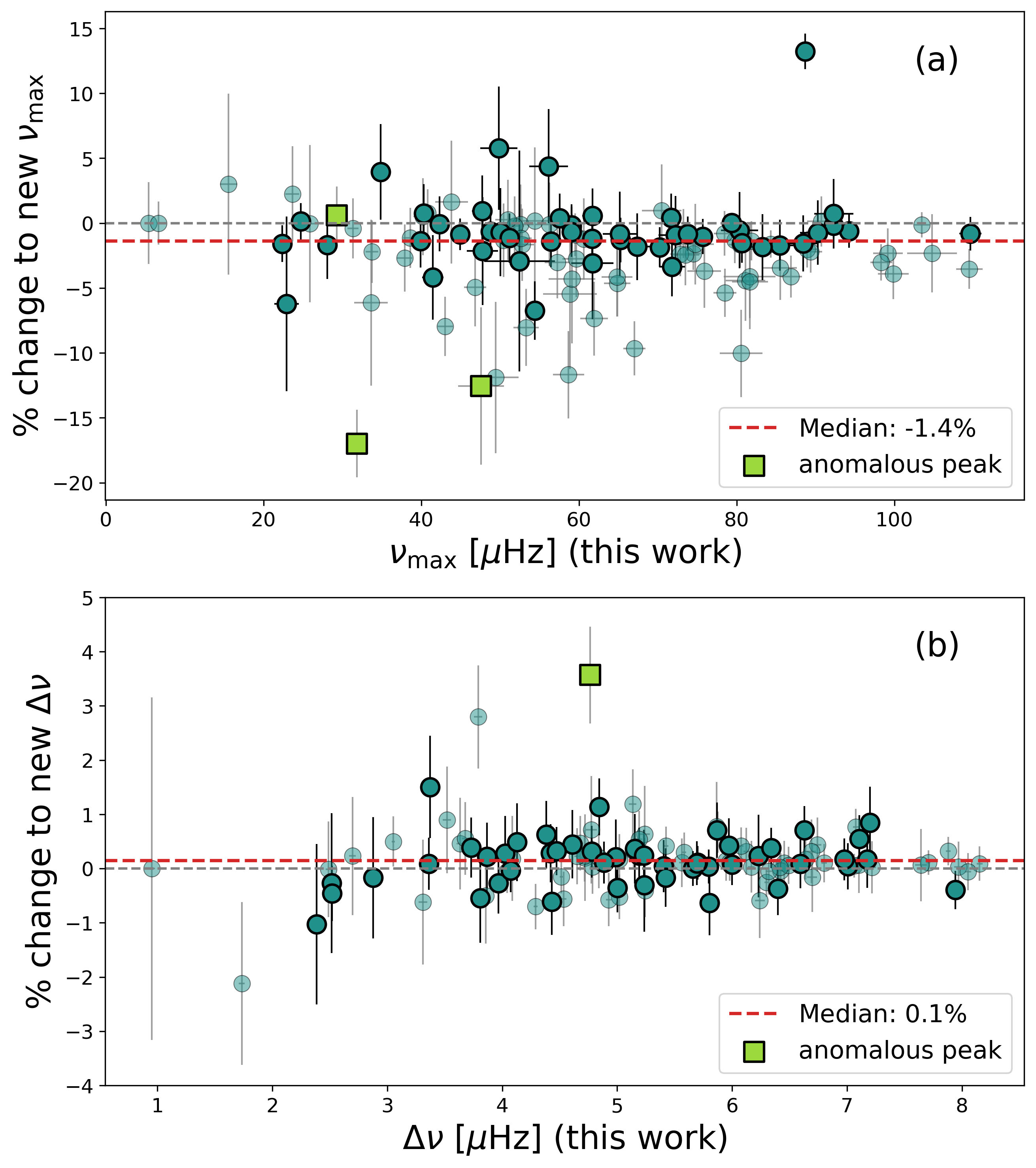}
    \caption{Two subplots showing the percentage changes in \numax{} (subplot a) and \Dnu{} (subplot b) from \yucat{} to this work, where negative values indicate a lower value in our work and positive values indicate a larger value in our work when comparing to \yucat{}. Both are plotted against the newly tabulated values for this work. Some of the horizontal error bars are too small to be visible. The darker blue points are those which are in the final sample of \Nofstars{} stars, and the lighter blue points are those which were analysed as part of the sample containing \totalNofstars{} stars. The three anomalous peak stars are denoted by green squares. In subplot b two off these anomalous peak stars lie off the axes, at percent differences of 14\% and 47\%. Each subplot also shows a horizontal line indicating the median of the percentage changes for each parameter.}
    \label{fig:fractional_change}
\end{figure}

\begin{figure*}
    \centering
    \includegraphics[width=\textwidth]{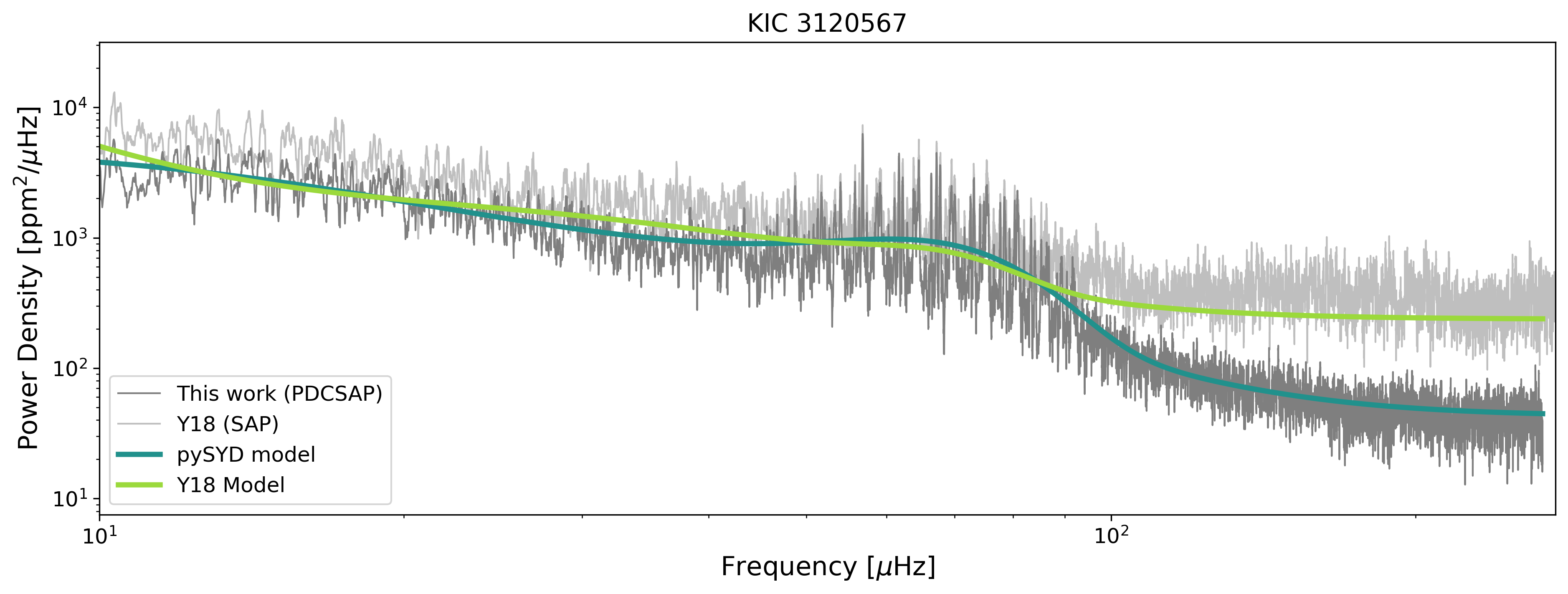}
    \caption{Here we show the power density spectrum and resulting background models from this work and \yucat{} for KIC 3120567, a memer of the final sample (M=3.19$\pm$0.32 \Msolar{}). We plot the power density spectrum with a boxcar smoothing (width of 0.02\Dnu{}) for this work (using the PDCSAP light curve) in dark grey. In light grey, we plot the similarly smoothed power density spectrum from \yucat{} (using the detrended SAP light curve). We additionally overplot the background models created from each of these power density spectra in blue and green for this work and \yucat{}, respectively. We point out the differences in these two models and power density spectra, especially in the white noise at high frequencies. We additionally draw attention to the difference in power excess amplitudes, as the differences in white noise between the two sources influence this measurement (see Section~\ref{subsec:envelopes}).}
    \label{fig:bgfit_compare}
\end{figure*}

\begin{figure}
    \centering
    \includegraphics[width=\columnwidth]{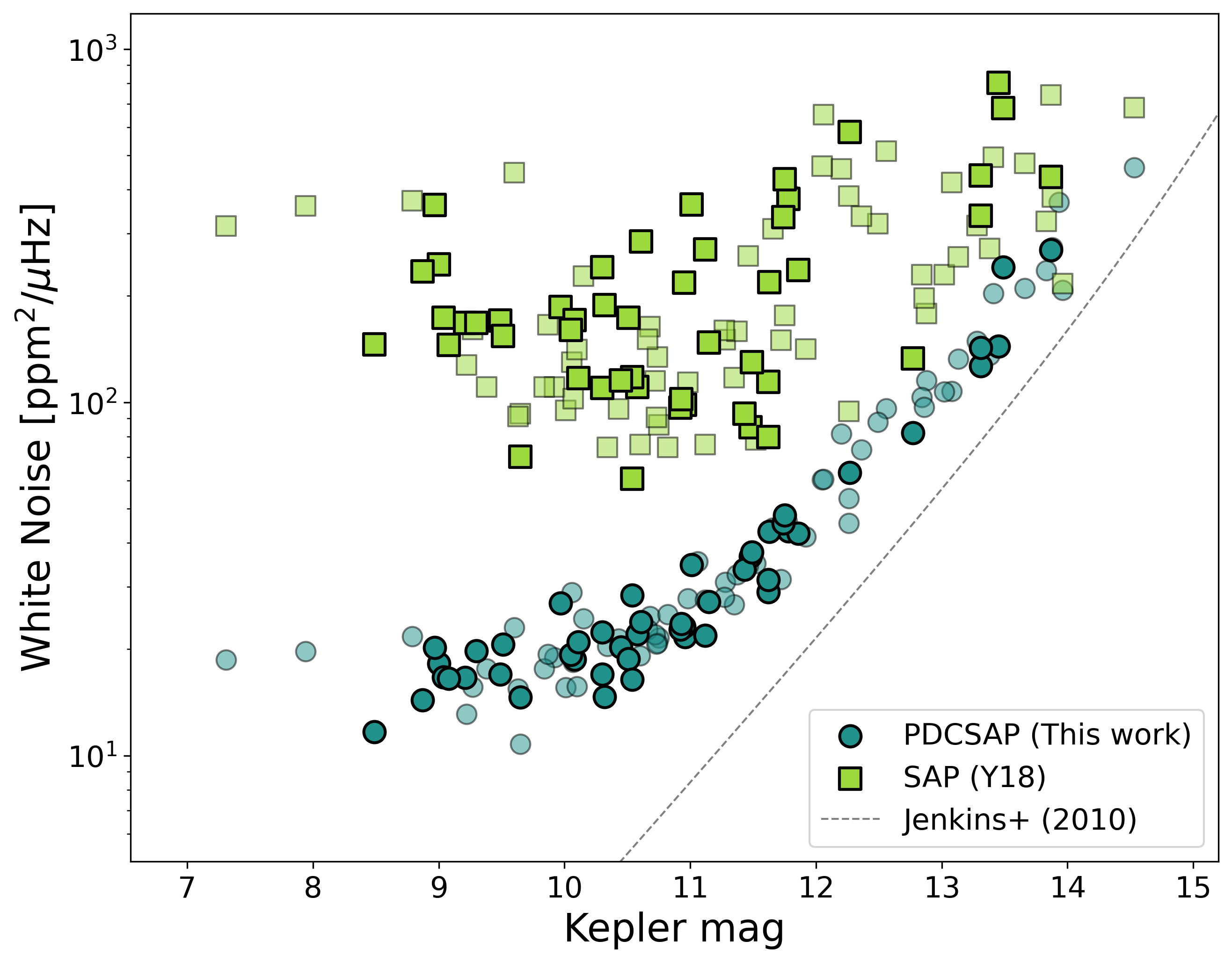}
    \caption{The white noise in the power spectrum as measured by \yucat{} using the SAP light curves (green) and this work using the PDCSAP light curves (blue) as a function of the Kepler magnitudes. The stars in our final sample of \Nofstars{} are denoted using opaque symbols, and the stars in the total sample of \totalNofstars{} stars are denoted by semi-transparent symbols. We additionally plot the theoretical minimum noise (photon noise) for Kepler 30-min data from \protect\citet{Jenkins2010_keplernoise} with a grey dashed line.}
    \label{fig:whitenoise}
\end{figure}

We present the percentage difference between the \numax{} measured in this work and those quoted in \yucat{} in Figure~\ref{fig:fractional_change}a. We see scatter in the measured \numax{} consistent with the quoted uncertainties, and a small systematic offset introduced by our refitting process.
The offset in \numax{} is due to differences in background noise levels between the two works. The background noise in red giants is made up of frequency-dependent granulation and white noise. Both {\sc pySYD} and {\sc SYD} begin by fitting this background. We compared our background fits from {\sc pySYD} and our power spectra to those generated for \yucat{} and found one key difference, illustrated by the example spectrum in Figure~\ref{fig:bgfit_compare}.
The white noise component of the background fit is typically an order of magnitude larger in \yucat{} than in our work (see Figure~\ref{fig:whitenoise}). This is due to the difference in the data processing on the light curves used by each work. \yucat{} uses detrended Kepler SAP light curves (see \yucat{} for description of this process), and this work uses the high-pass filtered PDCSAP light curves. 
We find that the white noise estimates for this work correlate strongly with Kepler magnitude whereas the white noise estimates for \yucat{} correlate only weakly with Kepler magnitude. We additionally show the theoretical lower limit of white noise (the photon noise) for Kepler 30-minute data as shown in \citet{Jenkins2010_keplernoise}, which neither \yucat{} nor this work reaches. We note that at Kepler magnitudes < 11, the white noise component of the fit is no longer dominated by photon noise, and therefore deviates from the \citet{Jenkins2010_keplernoise} relation.
It is clear that the light curves used in this work have lower overall white noise measurements, which we prefer for this analysis. This difference in white noise levels is what causes the small systematic decrease in \numax{} estimates. However, it is clear from Figure~\ref{fig:fractional_change} that the reduction in white noise does not affect estimation of \numax{} nor \Dnu{} very strongly, and the seismic parameters in \yucat{} are still reasonable estimates.

Additionally, we had three stars in our sample which we considered ``anamolous peak'' \citep{Colman2017_anamolouspeaks,Bedding2010_firstkepler} stars: KIC 6529078, KIC 3747623, and KIC 10384595. These stars exhibit a significantly large peak outside of the oscillation window, which may indicate the presence of a nearby close binary system or that the star is itself in a multiple system (see Section~\ref{subsec:binarity} for further discussion). 
Once the strong peak of these stars is removed, the \numax{} must be recalculated to the proper location. For each of these stars, the change in \numax{} from the original catalogue is large enough to remove the star from our final sample (see Figure~\ref{fig:fractional_change}).

\subsection{\Dnu{} adjustment}
\label{subsec:dnu}

There are a number of ways in which \Dnu{} can be measured. The {\sc SYD} pipeline and {\sc pySYD} both use an auto-correlation of a window around the power-excess. This generally works well, but can be affected by the mixed mode characteristic of the $\ell$=1 and $\ell$=2 modes. Other authors choose instead to use a weighted least-squares fit to the radial modes, where the weights are chosen to decrease with distance from \numax{}  \citep[e.g.,][]{White2011_solarlike_models,Sharma2016_fdnu,Stello2022_asfgridextension}. These works tend to include work with models, where one would have an exact value for the frequency, whereas for observations determining the exact pulsation frequency is less straightforward. There are many other ways to define and calculate \Dnu{}, for example \citet{Kallinger2012_epsilon} measures the separation between the three central radial orders, \citet{Mosser2011_universalpattern} measures the average separation for all observed p-modes, \citet{Campante2017_PSPS} uses the power spectrum of the power spectrum and, more recently, \citet{Dhanpal2022_MLseparations} uses machine learning techniques (this list is not exhaustive).
Additionally, \Dnu{} is not a constant as function of frequency. Indeed the definition of \Dnu{} itself, also known as the asymptotic relation:
\begin{equation}\label{eqn:asymptotic_relation}
\centering
\nu_{n,\ell} \approx \Delta \nu \left(n+\frac{\ell}{2}+\epsilon\right) + \delta \nu_{0,\ell}
\end{equation}
assumes that \Dnu{} is strictly independent of $n$. The deviation of \Dnu{} from the asymptotic relation can arise from abrupt changes, or glitches, in the interior stellar structure. It is known that higher mass red giants have stronger glitches than lower mass red giants, and red clump stars tend to also have glitches with stronger amplitudes \citep{Vrard2015_glitches}. When glitches are present, it is exceedingly difficult to define a precise \Dnu{}. However, when providing comparisons with other stars, \Dnu{} is still a useful estimate, as long as each star is measured in the same way. Many of the stars in our sample have p-mode glitches (see Section~\ref{subsec:glitches} for further discussion), however they tend to have similar shapes, and \Dnu{} is measured consistently across the sample. 


For our sample we used {\sc pySYD} to remeasure \Dnu{} for all stars. There is very little systematic difference between the \Dnu{} that we measure and the values quoted in \yucat, which is insignificant compared to the roughly 1\% scatter over the entire sample (see Figure~\ref{fig:fractional_change}b).

\subsection{T$_{\rm eff}$ adjustment}
\label{subsec:teff}

The temperatures quoted in \yucat{} were adopted from \citet{Mathur2017_temperatures}, which is a collection of stellar parameters from a variety of sources, explicitly citing estimates from spectroscopy for 14,813 stars, non-KIC photometry for 151,118 stars, and the KIC for 31,165 stars. While the method of collecting stellar parameters from various sources is good for studying a large sample of stars where the goal is coverage, for our smaller, more detailed sample we prefer to take all of our estimates from the same source if possible, to reduce systematic offsets.

We considered two large temperature estimation catalogues: APOGEE \citep{APOGEE_DR16} and LAMOST \citep{LAMOST}. The APOGEE and LAMOST temperatures show within 2\% agreement, with no systematic differences between the values estimated in each survey. It is not straightforward to place preference over either estimate when both are present in the same star. For this work, we adopt either the LAMOST or APOGEE temperatures if only one is available, otherwise we adopt the average of the two measurements. If neither of these temperatures are available, we adopt the temperature from \yucat. Only 13 stars are not present in LAMOST nor APOGEE, 7 of which remain in our final sample. The \Teff{} adoption source for every star is noted in our data set (see Table~\ref{tab:sample} notes). As we cannot provide homogenously measured \Teff{} values with only these two surveys, the \Teff{} remains a major source of potential uncertainty in our final mass estimates.

\subsection{Mass re-estimation and the final sample}
\label{subsec:finalsample}

These stars have been previously published in \yucat{} and thus have previously estimated asteroseismic masses. However, those asteroseismic masses depend on derived global oscillation parameters as follows in Eqn~\ref{eqn:mass_scaling}.
\citep{Ulrich1986_dnuscaling,Christensen-Dalsgaard1988_scalingrelations,Brown1991_numaxscaling,Kjeldsen1995_scalingrelations, Stello2008_wire, Kallinger2010_corot}:
\begin{equation}\label{eqn:mass_scaling}
\centering
\frac{M}{M_{\odot}} = \left(\frac{\nu_{\rm max}}{f_{\nu_{\rm max}}\nu_{\rm max,\odot}}\right)^{3} \left(\frac{\Delta \nu}{f_{\Delta \nu}\Delta \nu_{\odot}}\right)^{-4} \left(\frac{T_{\rm eff}}{T_{\rm eff,\odot}}\right)^{3/2}
\end{equation}
Thus, after re-calculating our estimates for \numax{}, \Dnu{}, and \Teff{}, we now re-estimate the masses for each of our stars.
For the solar values, we use $\nu_{\rm max, \odot}$ = 3090 $\pm$ 30 $\mu$Hz, $\Delta\nu_{\odot}$ = 135.1 $\pm$ 0.1 $\mu$Hz, and T$_{\rm eff, \odot}$ = 5772 K \citep{Huber2011_solarvalues, Prsa2016_solar_iau}. For the two correction factors, $f_{\Delta \nu}$ and $f_{\nu_{\rm max}}$, we used {\sc asfgrid} \citep{Sharma2016_fdnu,Stello2022_asfgridextension} to estimate $f_{\Delta \nu}$ (assuming our stars are core He-burning; see Section~\ref{subsec:evol_phase} for justification of this assumption) and assume $f_{\nu_{\rm max}}$ to be one, following the convention used by current works in red giant asteroseismology (such as \citealt{Yu2018_keplercatalog,Zinn2019_testingscalingrelation_gaia,Li2023_surfacecorrection}). {\sc asfgrid} requires input of \numax{}, \Teff{}, and log(g), the latter two of which we adopt from APOGEE and LAMOST (see Section~\ref{subsec:teff}). We note that the stellar model grid used in \citet{Sharma2016_fdnu} does not go beyond 4 \Msolar{}, however this upper mass limit was extended to 5.5 \Msolar{} in \citet{Stello2022_asfgridextension}. There are also more recent works such as \citet{Li2023_surfacecorrection} which calculate $f_{\Delta \nu}$ by taking into account the surface correction necessary in evolutionary models. These models span an even smaller range of masses, from 0.8 to 1.8 \Msolar{}, and we therefore cannot use the \citet{Li2023_surfacecorrection} models for our purposes. However, we note that the $f_{\Delta \nu}$ values found when including the surface correction terms tends to decrease $f_{\Delta \nu}$ by $\sim$1--2\%, which would decrease the mass estimated by roughly $\sim$4--8\%. 

\begin{figure}
    \centering
    \includegraphics[width=\columnwidth]{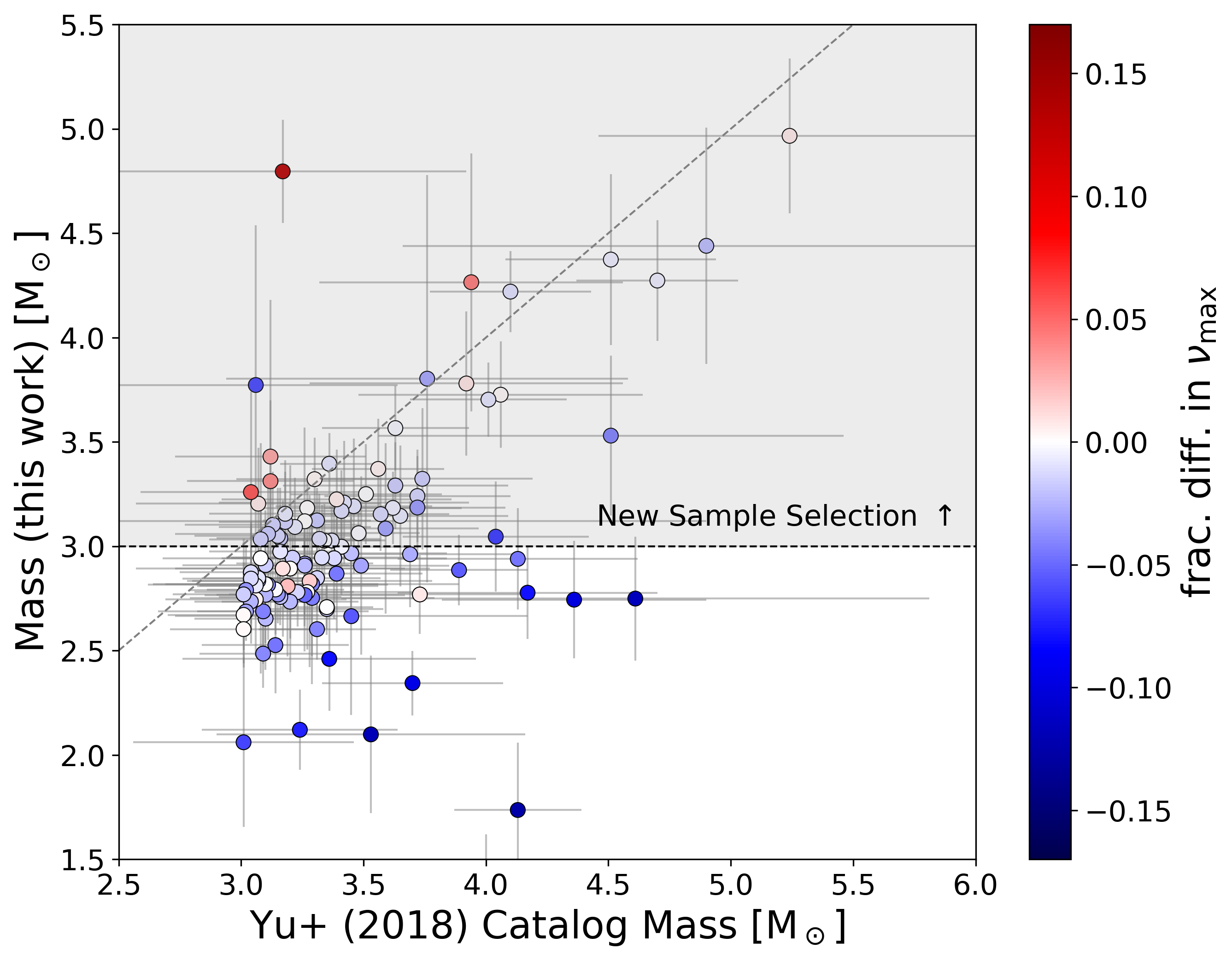}
    \caption{Plot of the masses generated for this work versus the \yucat{} catalogue masses, coloured by the fractional difference in \numax{}. Negative fractional differences indicate a lower \numax{} estimate than in \yucat{}. The one-to-one line is denoted by a grey dashed line. The black horizontal dashed line and the shaded region on the top denote the cutoff which defines the \Nofstars{} stars in the high-mass sample studied here.}
    \label{fig:mass_everything}
\end{figure}

Once we have validated the mass estimates for our stars, we again impose a cut in mass to 3.0 \Msolar{}, as the re-estimation process has decreased the mass estimates for the vast majority of stars.
We additionally remove any stars with \numax{} < 20 $\mu$Hz as those lower than this threshold have \Dnu{} values that are difficult to measure due to the small number of excited modes. This narrows our final sample to \Nofstars{} stars. The sample selection can be seen in Figure~\ref{fig:mass_everything}. While the re-estimation of \numax{}, \Dnu{}, and \Teff{} each effect the overall distribution of the masses (see the preceding subsections), it is clear that the systematic decrease in stellar mass from the \yucat{} catalogue is predominantly due to a systematic decrease in \numax{}. Section~\ref{subsec:numax} gives a full discussion of the origin of this offset. Our quoted uncertainties for the mass include the uncertainties on \numax{}, \Dnu{}, and \Teff{}, but we do not include systematic uncertainties such as the potential breakdown of the scaling relation at high mass. Future studies will reduce this uncertainty and refine the mass estimates.

\section{Sample demographics}
\label{sec:demographics}

\begin{table*}
	\centering
	\caption{The high-mass sample \label{tab:sample}}
	\begin{tabular}{cccccccccc} 
		KIC & Mass [\Msolar{}] & $\nu_{\rm max}$ [$\mu$Hz] & $\Delta\nu$ [$\mu$Hz] & Adopted $T_{\rm eff}$ [K] & $T_{\rm eff}$ source & $\epsilon$ & $\Delta\Pi$ [s] & $V_{\ell=1}^2$ & $V_{\ell=2}^2$\\
		\hline
            3347458 & 4.97 $\pm$ 0.37 & 40.29 $\pm$ 0.90 & 3.363 $\pm$ 0.016 & 4863 $\pm$ 84 & APOGEE & 0.89 $\pm$ 0.06 & --- & 0.81 $\pm$ 0.08 & 0.89 $\pm$ 0.08 \\
            9266192 & 4.80 $\pm$ 0.25 & 88.66 $\pm$ 1.21 & 6.341 $\pm$ 0.023 & 5193 $\pm$ 96 & APOGEE & 1.03 $\pm$ 0.05 & --- & 1.01 $\pm$ 0.09 & 0.55 $\pm$ 0.05 \\
            8378545 & 4.44 $\pm$ 0.57 & 47.77 $\pm$ 1.99 & 3.966 $\pm$ 0.022 & 4984 $\pm$ 44 & Average & 0.87 $\pm$ 0.07 & --- & 0.69 $\pm$ 0.09 & 0.42 $\pm$ 0.05 \\
            4756133 & 4.37 $\pm$ 0.41 & 80.36 $\pm$ 2.36 & 5.999 $\pm$ 0.023 & 5186 $\pm$ 95 & APOGEE & 0.93 $\pm$ 0.05 & --- & 0.70 $\pm$ 0.08 & 0.61 $\pm$ 0.04 \\
            5978324 & 4.27 $\pm$ 0.29 & 48.78 $\pm$ 0.97 & 4.073 $\pm$ 0.016 & 5051 $\pm$ 92 & APOGEE & 0.86 $\pm$ 0.05 & --- & 0.67 $\pm$ 0.05 & 0.48 $\pm$ 0.05 \\
            11518639 & 4.27 $\pm$ 0.62 & 56.15 $\pm$ 2.47 & 4.611 $\pm$ 0.029 & 5198 $\pm$ 187 & Y18 & 0.88 $\pm$ 0.08 & --- & 0.68 $\pm$ 0.14 & 0.52 $\pm$ 0.05 \\
            6599955 & 4.22 $\pm$ 0.19 & 75.71 $\pm$ 1.03 & 5.870 $\pm$ 0.030 & 5361 $\pm$ 27 & LAMOST & 0.94 $\pm$ 0.07 & --- & 0.53 $\pm$ 0.05 & 0.64 $\pm$ 0.06 \\
            9612933 & 3.80 $\pm$ 0.98 & 52.47 $\pm$ 4.46 & 4.433 $\pm$ 0.027 & 5096 $\pm$ 47 & Average & 0.83 $\pm$ 0.07 & --- & 0.57 $\pm$ 0.10 & 0.52 $\pm$ 0.03 \\
            7988900 & 3.78 $\pm$ 0.34 & 47.74 $\pm$ 1.31 & 4.128 $\pm$ 0.030 & 5088 $\pm$ 93 & APOGEE & 0.87 $\pm$ 0.08 & 358 $\pm$ 10 & 0.97 $\pm$ 0.12 & 0.60 $\pm$ 0.05 \\
            6382830 & 3.77 $\pm$ 0.77 & 22.93 $\pm$ 1.54 & 2.518 $\pm$ 0.013 & 4674 $\pm$ 41 & Average & 0.83 $\pm$ 0.05 & --- & 0.80 $\pm$ 0.11 & 0.68 $\pm$ 0.07 \\
            3955502 & 3.73 $\pm$ 0.25 & 24.71 $\pm$ 0.35 & 2.516 $\pm$ 0.032 & 5052 $\pm$ 51 & Average & 0.86 $\pm$ 0.13 & --- & 0.49 $\pm$ 0.15 & 0.40 $\pm$ 0.07 \\
            8569885 & 3.70 $\pm$ 0.18 & 44.93 $\pm$ 0.55 & 4.024 $\pm$ 0.028 & 5210 $\pm$ 48 & Average & 0.84 $\pm$ 0.08 & 302 $\pm$ 10 & 0.90 $\pm$ 0.11 & 0.63 $\pm$ 0.05 \\
            5097690 & 3.57 $\pm$ 0.20 & 59.04 $\pm$ 0.94 & 4.989 $\pm$ 0.035 & 5242 $\pm$ 49 & Average & 0.84 $\pm$ 0.08 & 316 $\pm$ 10 & 1.03 $\pm$ 0.18 & 0.73 $\pm$ 0.07 \\
            7175316 & 3.53 $\pm$ 0.38 & 41.49 $\pm$ 1.35 & 3.730 $\pm$ 0.021 & 5007 $\pm$ 142 & Y18 & 0.95 $\pm$ 0.06 & 250 $\pm$ 10 & 1.08 $\pm$ 0.13 & 0.78 $\pm$ 0.05 \\
            8230626 & 3.40 $\pm$ 0.15 & 109.62 $\pm$ 1.41 & 7.942 $\pm$ 0.028 & 5150 $\pm$ 47 & Average & 1.10 $\pm$ 0.05 & 198 $\pm$ 10 & 1.56 $\pm$ 0.09 & 0.67 $\pm$ 0.04 \\
            8525150 & 3.37 $\pm$ 0.24 & 71.69 $\pm$ 1.32 & 5.660 $\pm$ 0.018 & 5002 $\pm$ 142 & Y18 & 0.97 $\pm$ 0.04 & 234 $\pm$ 10 & 0.99 $\pm$ 0.14 & 0.62 $\pm$ 0.06 \\
            7971558 & 3.32 $\pm$ 0.34 & 28.11 $\pm$ 0.74 & 2.877 $\pm$ 0.032 & 5176 $\pm$ 164 & Y18 & 0.81 $\pm$ 0.11 & 351 $\pm$ 10 & 0.39 $\pm$ 0.08 & 0.43 $\pm$ 0.07 \\
            9468199 & 3.32 $\pm$ 0.20 & 57.54 $\pm$ 1.09 & 4.886 $\pm$ 0.019 & 5097 $\pm$ 47 & Average & 0.91 $\pm$ 0.05 & 326 $\pm$ 10 & 1.13 $\pm$ 0.22 & 0.71 $\pm$ 0.07 \\
            10621713 & 3.31 $\pm$ 0.39 & 34.87 $\pm$ 1.28 & 3.373 $\pm$ 0.032 & 5143 $\pm$ 34 & LAMOST & 0.78 $\pm$ 0.10 & 268 $\pm$ 10 & 1.05 $\pm$ 0.23 & 0.74 $\pm$ 0.28 \\
            9286851 & 3.29 $\pm$ 0.21 & 85.49 $\pm$ 1.67 & 6.627 $\pm$ 0.029 & 5141 $\pm$ 46 & Average & 0.97 $\pm$ 0.06 & 225 $\pm$ 10 & 1.54 $\pm$ 0.10 & 0.69 $\pm$ 0.05 \\
            11045134 & 3.26 $\pm$ 0.48 & 49.84 $\pm$ 2.37 & 4.384 $\pm$ 0.027 & 5061 $\pm$ 105 & APOGEE & 0.87 $\pm$ 0.07 & --- & 0.43 $\pm$ 0.16 & 0.54 $\pm$ 0.05 \\
            9245283 & 3.25 $\pm$ 0.24 & 42.31 $\pm$ 0.90 & 3.866 $\pm$ 0.024 & 5039 $\pm$ 91 & APOGEE & 0.78 $\pm$ 0.07 & 334 $\pm$ 10 & 0.94 $\pm$ 0.10 & 0.61 $\pm$ 0.05 \\
            10094550 & 3.24 $\pm$ 0.22 & 56.47 $\pm$ 1.11 & 4.844 $\pm$ 0.025 & 5099 $\pm$ 92 & APOGEE & 0.88 $\pm$ 0.06 & 273 $\pm$ 10 & 0.95 $\pm$ 0.09 & 0.68 $\pm$ 0.06 \\
            4348593 & 3.22 $\pm$ 0.24 & 61.70 $\pm$ 1.31 & 5.155 $\pm$ 0.033 & 5058 $\pm$ 92 & APOGEE & 0.92 $\pm$ 0.08 & 272 $\pm$ 10 & 1.33 $\pm$ 0.08 & 0.54 $\pm$ 0.04 \\
            4940439 & 3.21 $\pm$ 0.29 & 72.24 $\pm$ 2.16 & 5.798 $\pm$ 0.018 & 5038 $\pm$ 46 & Average & 0.98 $\pm$ 0.04 & 236 $\pm$ 10 & 0.69 $\pm$ 0.08 & 0.55 $\pm$ 0.05 \\
            2845610 & 3.20 $\pm$ 0.27 & 92.31 $\pm$ 2.47 & 7.108 $\pm$ 0.031 & 5207 $\pm$ 48 & Average & 1.00 $\pm$ 0.06 & --- & 0.81 $\pm$ 0.09 & 0.76 $\pm$ 0.04 \\
            3120567 & 3.19 $\pm$ 0.32 & 65.17 $\pm$ 2.13 & 5.420 $\pm$ 0.029 & 5099 $\pm$ 45 & Average & 0.93 $\pm$ 0.06 & 220 $\pm$ 10 & 1.68 $\pm$ 0.08 & 0.80 $\pm$ 0.04 \\
            10736390 & 3.19 $\pm$ 0.25 & 71.81 $\pm$ 1.63 & 5.802 $\pm$ 0.035 & 5068 $\pm$ 91 & APOGEE & 0.88 $\pm$ 0.07 & 249 $\pm$ 10 & 1.88 $\pm$ 0.31 & 1.06 $\pm$ 0.07 \\
            6866251 & 3.18 $\pm$ 0.17 & 94.23 $\pm$ 1.21 & 7.200 $\pm$ 0.047 & 5174 $\pm$ 95 & APOGEE & 0.92 $\pm$ 0.09 & 250 $\pm$ 10 & 1.35 $\pm$ 0.14 & 0.60 $\pm$ 0.05 \\
            4372082 & 3.18 $\pm$ 0.13 & 79.38 $\pm$ 0.55 & 6.229 $\pm$ 0.047 & 5049 $\pm$ 50 & Average & 0.94 $\pm$ 0.10 & 206 $\pm$ 10 & 1.34 $\pm$ 0.08 & 0.65 $\pm$ 0.04 \\
            5307930 & 3.17 $\pm$ 0.19 & 51.21 $\pm$ 0.97 & 4.472 $\pm$ 0.020 & 4990 $\pm$ 45 & Average & 0.91 $\pm$ 0.05 & 316 $\pm$ 10 & 0.93 $\pm$ 0.11 & 0.65 $\pm$ 0.07 \\
            11456735 & 3.16 $\pm$ 0.26 & 90.28 $\pm$ 2.24 & 7.008 $\pm$ 0.030 & 5210 $\pm$ 98 & APOGEE & 1.05 $\pm$ 0.05 & 236 $\pm$ 10 & 1.26 $\pm$ 0.11 & 0.63 $\pm$ 0.06 \\
            4562675 & 3.15 $\pm$ 0.18 & 65.28 $\pm$ 1.13 & 5.403 $\pm$ 0.026 & 5064 $\pm$ 31 & LAMOST & 0.91 $\pm$ 0.06 & 266 $\pm$ 10 & 1.19 $\pm$ 0.14 & 0.69 $\pm$ 0.05 \\
            4370592 & 3.15 $\pm$ 0.34 & 50.03 $\pm$ 1.70 & 4.420 $\pm$ 0.024 & 5046 $\pm$ 91 & APOGEE & 0.92 $\pm$ 0.06 & 308 $\pm$ 10 & 1.26 $\pm$ 0.14 & 0.56 $\pm$ 0.07 \\
            4273491 & 3.12 $\pm$ 0.16 & 70.23 $\pm$ 1.08 & 5.700 $\pm$ 0.018 & 5050 $\pm$ 52 & Average & 1.03 $\pm$ 0.04 & --- & 0.81 $\pm$ 0.09 & 0.72 $\pm$ 0.05 \\
            9786910 & 3.12 $\pm$ 0.24 & 22.36 $\pm$ 0.32 & 2.384 $\pm$ 0.035 & 4892 $\pm$ 85 & APOGEE & 0.68 $\pm$ 0.14 & --- & 1.34 $\pm$ 0.14 & 1.00 $\pm$ 0.07 \\
            12020628 & 3.10 $\pm$ 0.25 & 88.41 $\pm$ 1.91 & 6.978 $\pm$ 0.027 & 5239 $\pm$ 166 & Y18 & 1.07 $\pm$ 0.05 & 201 $\pm$ 10 & 1.62 $\pm$ 0.15 & 0.85 $\pm$ 0.06 \\
            10809272 & 3.09 $\pm$ 0.24 & 59.06 $\pm$ 1.13 & 5.232 $\pm$ 0.025 & 5403 $\pm$ 168 & Y18 & 0.98 $\pm$ 0.05 & 319 $\pm$ 10 & 0.95 $\pm$ 0.08 & 0.61 $\pm$ 0.05 \\
            5106376 & 3.08 $\pm$ 0.41 & 61.69 $\pm$ 2.67 & 5.191 $\pm$ 0.026 & 5042 $\pm$ 46 & Average & 0.94 $\pm$ 0.06 & --- & 0.70 $\pm$ 0.08 & 0.63 $\pm$ 0.04 \\
            10322513 & 3.06 $\pm$ 0.15 & 92.30 $\pm$ 0.92 & 7.174 $\pm$ 0.038 & 5211 $\pm$ 108 & APOGEE & 1.10 $\pm$ 0.07 & 246 $\pm$ 10 & 1.14 $\pm$ 0.08 & 0.69 $\pm$ 0.06 \\
            8395466 & 3.06 $\pm$ 0.26 & 67.37 $\pm$ 1.73 & 5.695 $\pm$ 0.023 & 5250 $\pm$ 99 & APOGEE & 0.91 $\pm$ 0.05 & --- & 0.49 $\pm$ 0.07 & 0.72 $\pm$ 0.05 \\
            4940935 & 3.05 $\pm$ 0.23 & 39.94 $\pm$ 0.82 & 3.810 $\pm$ 0.031 & 5180 $\pm$ 98 & APOGEE & 0.94 $\pm$ 0.09 & 335 $\pm$ 10 & 0.77 $\pm$ 0.07 & 0.69 $\pm$ 0.07 \\
            8037930 & 3.05 $\pm$ 0.26 & 54.42 $\pm$ 1.22 & 4.777 $\pm$ 0.023 & 5084 $\pm$ 173 & Y18 & 0.97 $\pm$ 0.05 & 288 $\pm$ 10 & 1.21 $\pm$ 0.08 & 0.61 $\pm$ 0.06 \\
            7581399 & 3.04 $\pm$ 0.24 & 83.28 $\pm$ 2.11 & 6.595 $\pm$ 0.030 & 5135 $\pm$ 53 & Average & 1.05 $\pm$ 0.06 & 220 $\pm$ 10 & 1.38 $\pm$ 0.11 & 0.84 $\pm$ 0.05 \\
            11044315 & 3.04 $\pm$ 0.22 & 61.68 $\pm$ 1.14 & 5.237 $\pm$ 0.045 & 5095 $\pm$ 94 & APOGEE & 0.97 $\pm$ 0.10 & 327 $\pm$ 10 & 1.20 $\pm$ 0.10 & 0.78 $\pm$ 0.05 \\
            5707338 & 3.03 $\pm$ 0.16 & 80.65 $\pm$ 1.24 & 6.397 $\pm$ 0.031 & 5063 $\pm$ 45 & Average & 1.05 $\pm$ 0.06 & 244 $\pm$ 10 & 1.46 $\pm$ 0.14 & 0.84 $\pm$ 0.09 \\
            11235672 & 3.03 $\pm$ 0.16 & 73.81 $\pm$ 0.97 & 5.971 $\pm$ 0.030 & 5051 $\pm$ 91 & APOGEE & 1.00 $\pm$ 0.06 & 270 $\pm$ 10 & 1.23 $\pm$ 0.10 & 0.56 $\pm$ 0.05 \\
            11413158 & 3.03 $\pm$ 0.18 & 59.05 $\pm$ 0.95 & 5.003 $\pm$ 0.022 & 4976 $\pm$ 89 & APOGEE & 0.97 $\pm$ 0.05 & 211 $\pm$ 10 & 1.07 $\pm$ 0.11 & 0.59 $\pm$ 0.06 \\
		\hline
            \multicolumn{9}{l}{Note: ``Average" in the $T_{\rm eff}$ source column refers to an average of the APOGEE and LAMOST $T_{\rm eff}$ values.} \\
            \multicolumn{9}{l}{Note: An extended version of this table is available online. This version contains columns of contamination/binarity factors (our contamination } \\
            \multicolumn{9}{l}{flag, Kepler CROWDSAP metric, Gaia RUWE, See Section~\ref{subsec:binarity}).} \\ 
	\end{tabular}
\end{table*}

\begin{figure}
    \centering
    \includegraphics[width=\columnwidth]{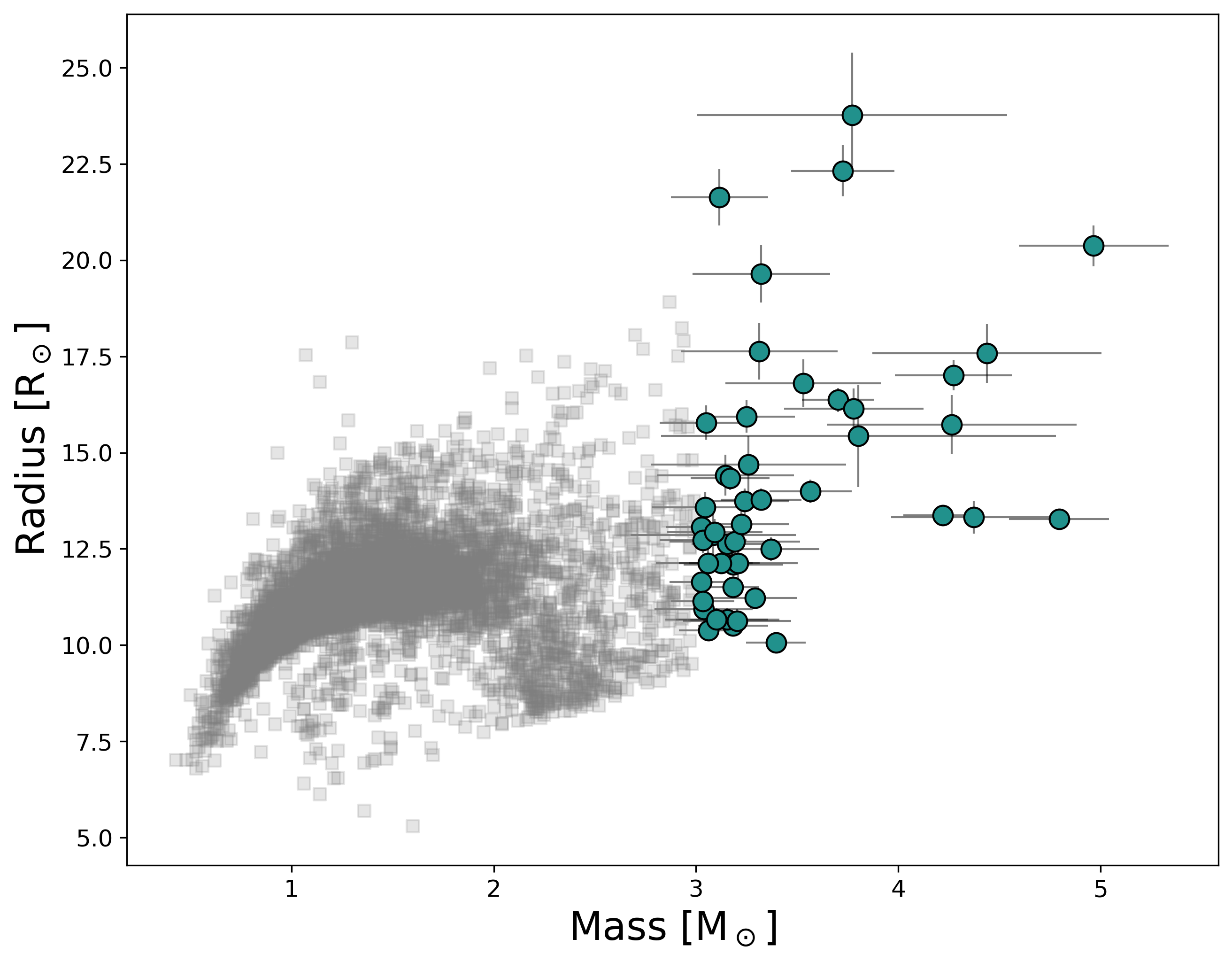}
    \caption{A radius versus mass plot showing our sample (blue circles) with all stars that were identified as clump stars in \yucat{} (grey squares).}
    \label{fig:mass_radius}
\end{figure}

We present our final high-mass sample in Table~\ref{tab:sample}. Most of the stars in this sample are very close to the lower mass boundary of 3 \Msolar{}, in fact 28 out of the 48 stars lie within 3--3.25 \Msolar{}, with 6 stars from 3.25--3.5 \Msolar{}, 7 stars from 3.5--4.0 \Msolar{}, and 7 stars with M$>$4 \Msolar{}. The stars cover a range in \numax{} and \Dnu{}, however we do note that our entire sample shows \numax{} values less than 110 $\mu$Hz. We show the mass and radius data for our sample in Figure~\ref{fig:mass_radius} in comparison with all stars labelled as clump stars in \yucat{}.

\begin{figure*}
    \centering
    \includegraphics[width=\textwidth]{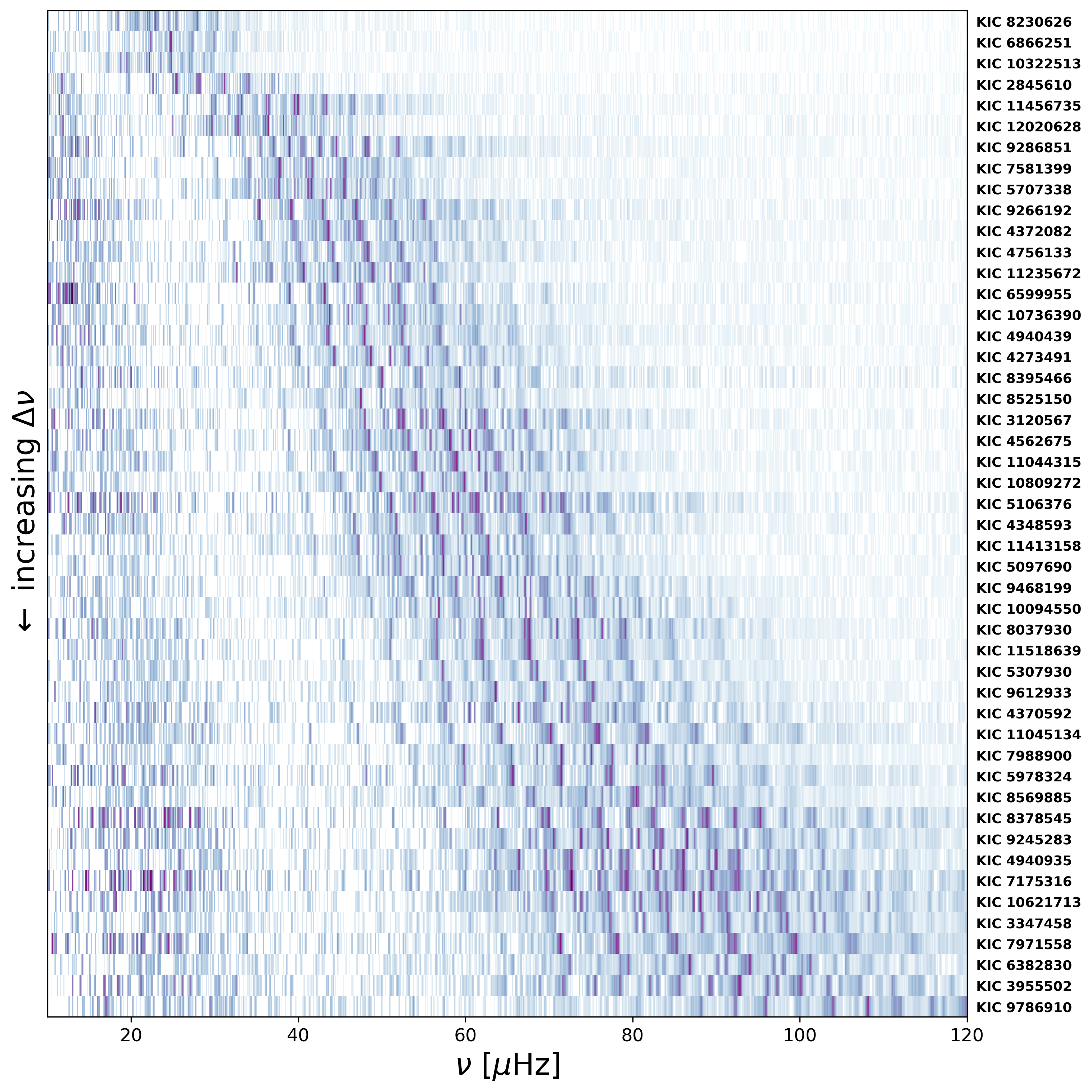}
    \caption{Power spectra for the \Nofstars{} stars in our high-mass sample. Each row represents a different star, sorted with \Dnu{} increasing downwards. Each star is labelled with its KIC number to the right. The greyscale represents the power as a function of frequency. Note that the power excess associated with solar-like oscillations widens with increasing \Dnu{} and \numax.}
    \label{fig:stacked_spectra}
\end{figure*}

We show the background-subtracted oscillation spectra of all our stars in Figure~\ref{fig:stacked_spectra}, ordered by increasing \Dnu{} (which also corresponds to increasing \numax, given the narrow mass range). Here one can easily see the relationship between the power excess width and \numax{}, which will be further discussed in Section~\ref{subsec:envelopes}. The lower-frequency ends of these spectra shows some high amplitude noise left over from the granulation background subtraction. One can also see the range of dipole mode visibilities from this figure (see Section~\ref{subsec:suppression} for further discussion). 

\begin{figure}
    \centering
    \includegraphics[width=\columnwidth]{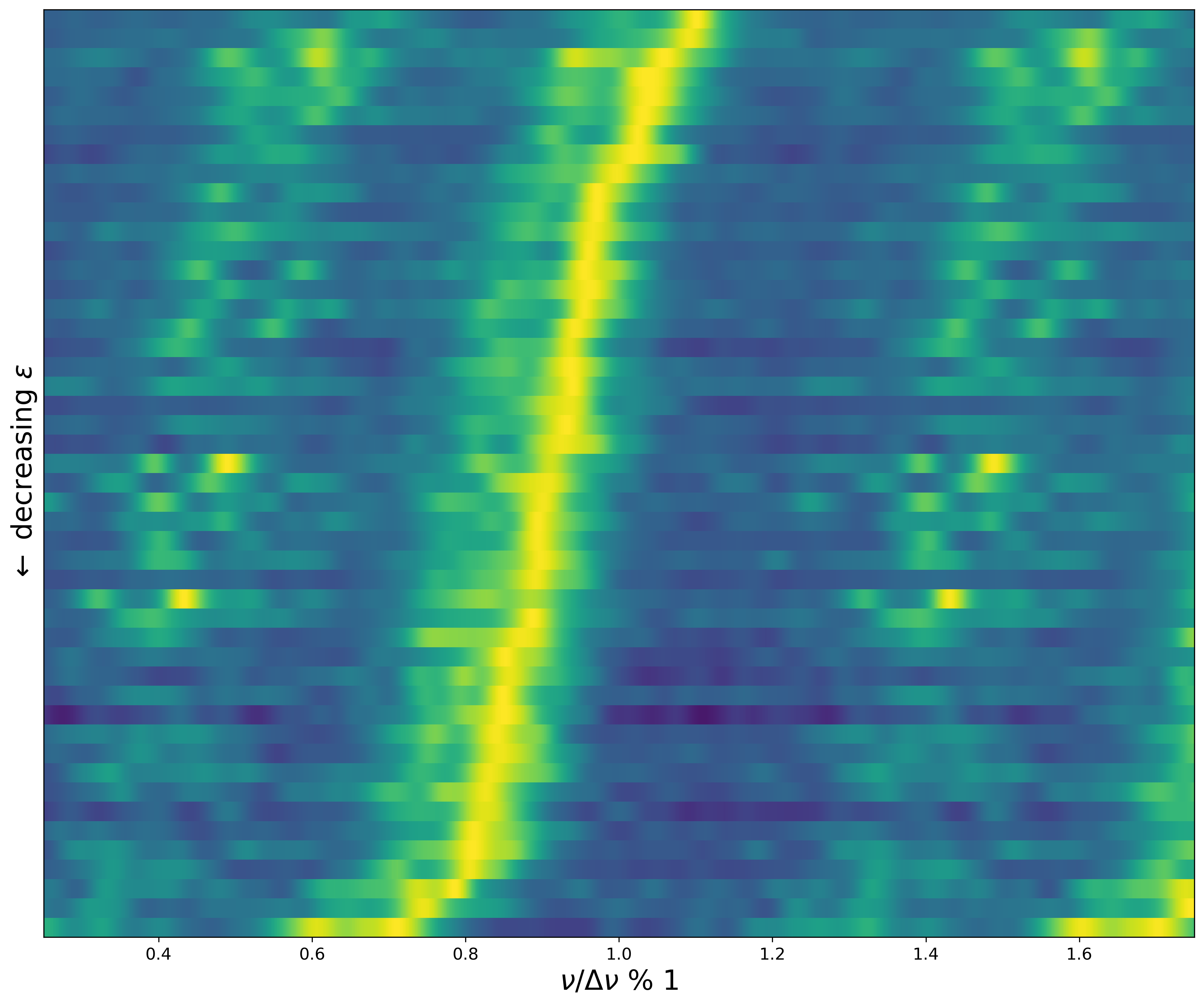}
    \caption{A representation of the collapsed frequency \'echelles for our sample. Each row represents a different star, sorted by the phase term $\epsilon$. The colour map denotes the power plotted as a function of frequency divided by \Dnu{} modulo one. The collapsed \'echelle has been replicated such that the radial mode ridge wraps around $\epsilon$=1. The bright yellow strip at roughly $\nu$/$\Dnu \mod 1$ of 0.8--1 denotes the location of the radial modes, and thus gives an estimate of the phase shift, $\epsilon$. Note that $\epsilon$ is defined to be between $\sim$0.5-1.5. The weak vertical strip to the left of the radial mode ridge (the brightest one) denotes the quadrupole ridge, and the broad ridges outside of these represent the dipole modes.}
    \label{fig:stacked_epsilon}
\end{figure}

We also compute collapsed frequency \'echelles, which are presented in Figure~\ref{fig:stacked_epsilon}. These are ordered by decreasing phase shift, $\epsilon$. as measured from the location of the bright vertical ridge corresponding to the radial modes. Most of our stars have $\epsilon$ between 0.8 and 1.0.
To the left of the radial ridge one can see the quadrupole ridge faintly, and further to the left (around $\nu$/$\Dnu$\%1 $\sim$ 0.5) lies the dipole ridge. It seems that the stars with smallest $\epsilon$ tend to have more suppressed dipole modes (See Section~\ref{subsec:suppression} for more discussion).


\subsection{Binarity and Contamination}
\label{subsec:binarity}

High-mass red giants are an interesting laboratory for studying binary systems. Binary fraction is known to increase with stellar mass, with some current estimates finding that up to 50 to 70 \% of young A- and late B-type stars (the progenitors of the high-mass red giants) are in binaries (see review by \citealt{Lee2020_binaryfraction}). It is also known that binarity, especially close binarity, weakens oscillation amplitudes in red giants \citep{Gaulme2014_binarysuppression,Schonhut-Stasik2020_binarysuppression}. A review of binary orbital period as a function of primary mass and mass ratio can be found in \citet{Moe2017_multiplicity}. We therefore check if our sample has any indications of binarity, and additionally check for the likelihood of contamination.

Yu+18 and the catalogues from which those data were collated removed any obvious binary signals (i.e. eclipsing systems), but there may still be signals hidden within. An excellent example of these is the class of Kepler stars with ``anomalous peaks" which may be due to nearby (chance-aligned) or associated close binaries (in hierarchical triples) \citep{Colman2017_anamolouspeaks}. As discussed in Section~\ref{subsec:numax}, our original set of \totalNofstars{} contained three such stars: KIC 6529078 and KIC 3747623, which were previously known, and KIC 10384595, which was not previously published.
These signals are understood to come either from chance alignments with other systems, or from physically associated systems such as hierarchical triples. 
As mentioned in Section~\ref{subsec:numax}, these stars with anomalous peaks are not present in the final sample.

We attempt to quantify the probability that binarity or contamination are occurring through various metrics, which we record in our extended data table (available online). The first of these is through visual inspection of the Kepler TPFs and the pipeline apertures. Using the {\sc lightkurve} \citep{lightkurve} package function \texttt{interact\_sky}, one can view the Gaia DR2 \citep{Gaia_DR2} positions overlain on the Kepler Target Pixel File (TPF). We used this to view whether there were any known sources nearby the target star, specifically within the Kepler aperture. For each star, we added a flag in Table~\ref{tab:sample} to indicate whether the star has (i)~no nearby potential contaminating sources (denoted with 0), (ii)~nearby sources that were more than 3 mag fainter than the target star, and thus not expected to be strongly contaminating the signal (denoted with 2), or (iii)~a nearby star of similar brightness (denoted with 1). 
For the final high-mass sample, these three scenarios represented 44\%, 46\%, and 10\% of the sample, respectively. Note that we were quite strict with the definition of ``potentially contaminated" as meaning that there was a star within the TPF of a brightness within 3 mag of the target. Interestingly, none of the three anomalous peak stars have potentially contaminating nearby stars that have been catalogued by Gaia DR2. The package we use for this study (the {\sc lightkurve} function \texttt{interact\_sky}) does not allow us to compare with the Gaia DR3 positions, but we do not expect the update from DR2 to DR3 to shift targets enough to affect this analysis.

The Kepler database quotes a crowding metric (\texttt{CROWDSAP}), which is defined as the ratio of the target flux to the total flux within the aperture. We find only two stars in the final sample with values < 0.95 (three in the large sample), only one of which has a potentially contaminating source. The other star has no nearby sources resolved by Gaia DR2. We include this value in Table~\ref{tab:sample}, for completeness. We also include Gaia's renormalized unit weight error (RUWE) parameter, which may indicate binarity for values higher than 1.4 \citep{Fabricius2021_gaiaedr3validation}. The large sample has 15 stars with RUWE > 1.4, whereas the final sample has seven. The final sample has no star with RUWE > 3.5, however the large sample has three, with the maximum in the large sample being KIC 5380617 which has RUWE = 43.331. Of the three anomalous peak stars, only KIC 3747623 has a RUWE value above the threshold, with RUWE = 11.013. We emphasize that a low RUWE value does not exclude the possibility of binarity, nor is a large RUWE value exclusive to binary systems.

We additionally find eleven stars in our large sample that have been suggested as potential spectroscopic binaries in three different articles. 
KIC 9655167 and KIC 9655101 were reported as single-lined spectroscopic binaries by \citet{MolendaZakowicz2014_NGC6811SB} in their study of the Kepler open cluster NGC 6811. These two stars are also listed as NGC 6811 cluster members by \citet{Stello2011_keplerclusters}. Note that this is a young cluster with high-mass stars currently in the red giant/clump phase, consistent with the mass estimates of these stars.
We find KIC 9836930 in the reference sample (Sample R) of \citet{Jorissen2020_SBpaper} as a potential single-lined spectroscopic binary.
Finally, stars with KIC IDs 8365782, 10094550, 6866251, 11297585, 5080332, 11413158, 5534910, and 5707338 appear in the Gaia DR3 non-single star catalogue \citep{GaiaDR3_nonsinglestarcatalog,GaiaDR3_release} as potential single-lined spectroscopic binaries, although only one of these (KIC 11413158) has a strong significance estimation.


To summarize, we cannot rule out the fact that some of our stars may either be in multiple systems or otherwise contaminated in some way. With the combination of the aforementioned metrics, we have an understanding of which stars are more likely to be contaminated or in binaries. That being said, we do not currently believe any of our stars are having strong interactions with a companion source, nor a nearby star contaminating the oscillation signals, as we do not see those signatures in the Kepler pixel data nor the oscillation spectra. In fact, we do not see any clear indications that any of the stars in our sample are in binaries.


\subsection{Evolutionary Phase}
\label{subsec:evol_phase}

Near the red giant branch (RGB) region of the HRD, there is significant overlap between stars of different evolutionary stages, specifically the RGB itself, the red clump (often called the horizontal branch in globular clusters), and the asymptotic giant branch (AGB). These three represent the evolutionary phases corresponding to shell hydrogen burning, core helium burning (CHeB), and shell hydrogen and helium burning\footnote{Note that there are a few differing definitions of the beginning of the AGB. For this work we define the beginning of the AGB to be at core He exhaustion and the onset of shell He burning.}, respectively. Additionally, at masses greater than $\sim$ 2 \Msolar{}, stars do not undergo explosive ignition of core-He burning (``the Helium Flash"), and therefore settle onto a region called the ``secondary clump", which lies separate from the lower-mass red clump \citep{Girardi1999_secondaryclump}. 
All of our stars have masses greater than 2 \Msolar{}, so if they are in the core He-burning phase, they are going to lie in the secondary clump. In the following text, we will refer to these three phases as RGB, CHeB, and AGB, and use the term ``red giants" as a general term that encompasses these three phases. For our sample, we explore the potential evolutionary phases for each star.

\begin{figure}
    \centering
    \includegraphics[width=\columnwidth]{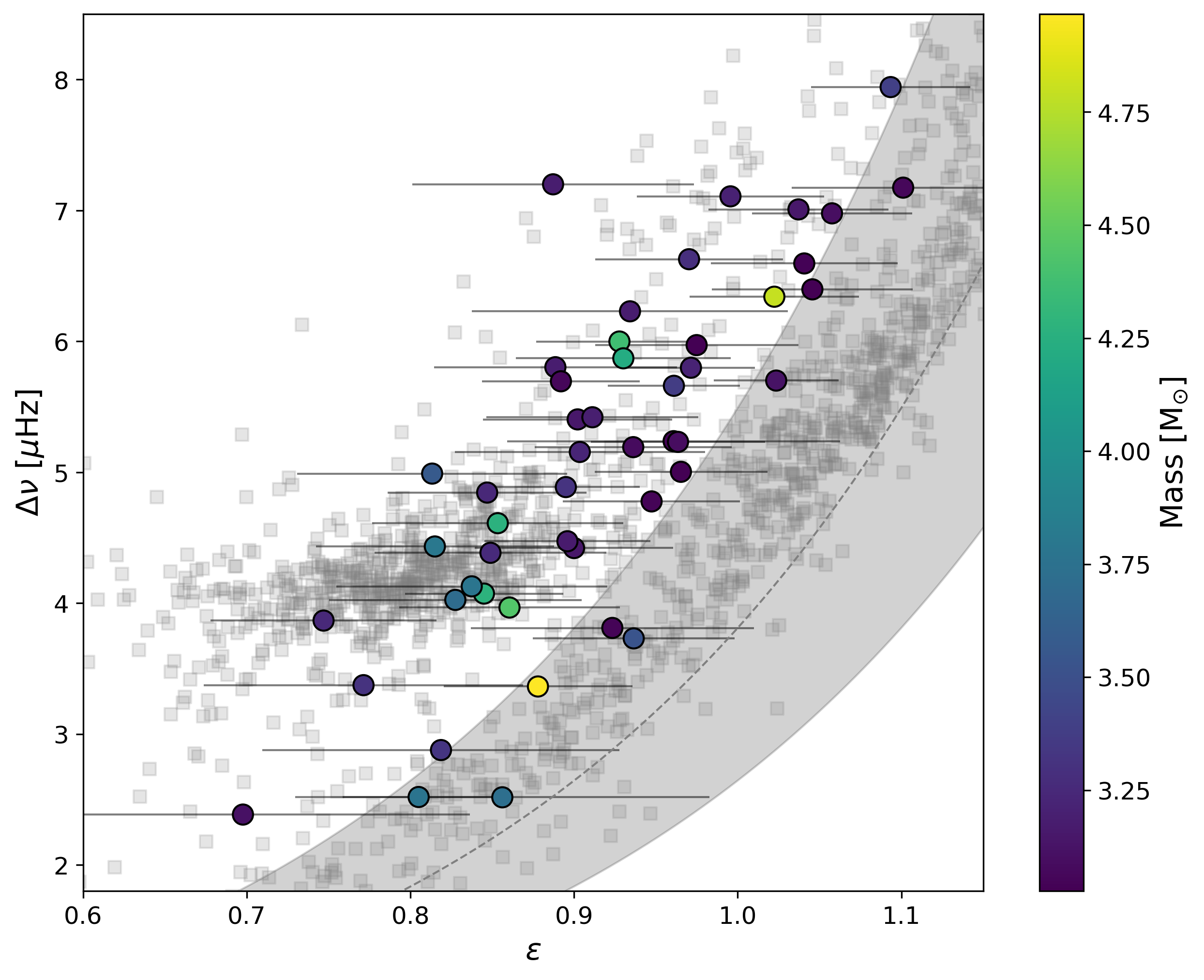}
    \caption{Here we show \Dnu{} as a function of the phase shift $\epsilon$, with the points coloured by the mass calculated within this work. 
    The error bars on \Dnu{} are smaller than the marker size.
    In the background, in grey squares, we plot the $\epsilon_c$ values for Kepler red giants estimated using the \protect\citet{Kallinger2012_epsilon} method from \citet{Kallinger_peakbaggingrepo}. The grey dashed line is the relation between \Dnu{} and $\epsilon$ derived by \protect\citet{Corsaro2012_epsilonrelation} for RGB stars in Kepler clusters, and the shaded region is $\pm$ 0.1 in $\epsilon$ from that relation, to mirror the selection in \protect\citet{Stello2016_visPASA}.}
    \label{fig:epsilon}
\end{figure}

We can first test the evolutionary phase using the value of $\epsilon$, or the position of the radial mode in units of \Dnu{} (sometimes called the phase offset), as defined by the asymptotic relation (Eqn~\ref{eqn:asymptotic_relation}). We measure this by calculating the location of the maximum value of a collapsed \'echelle (where the values for each radial order of a typical \'echelle diagram are summed, see Figure~\ref{fig:stacked_epsilon}). 
The uncertainty associated with $\epsilon$ is strongly tied to the uncertainty in \Dnu{} because changing \Dnu{} directly affects the measurement of $\epsilon$. We therefore adopt an uncertainty on $\epsilon$ equivalent to the fractional error on \Dnu{} multiplied by the approximate radial order near \numax{}, calculated as $\sigma_{\Delta\nu}(\nu_{\rm max}/\Delta\nu)$ for each star.
The value of $\epsilon$ differs for CHeB and RGB stars due to the change in the acoustic depth of the partial Helium ionization zone \citep{Bedding2011_nature,Kallinger2012_epsilon,Christensen-Dalsgaard2014_epsilon}. In Figure~\ref{fig:epsilon} we show the relation between \Dnu{} and $\epsilon$ for our sample compared to the central $\epsilon_c$ values from \citet{Kallinger_peakbaggingrepo} as measured using the method in \citet{Kallinger2012_epsilon}. The dashed line is a fitted relation from \citet{Corsaro2012_epsilonrelation} based on RGB stars in Kepler clusters. The shaded region denotes $\epsilon$ $\pm$ 0.1 from this relation, as was done in \citet{Stello2016_visPASA} to isolate the red giant branch stars. The majority of our stars lie clearly offset from this relation, indicating that they are likely in the CHeB phase \citep{Bedding2011_nature}. This agrees as well with the position of the secondary clump stars in \citet{Kallinger2012_epsilon}, as seen in the comparison sample. We however note that the works of \citet{Kallinger2012_epsilon} and \citet{Christensen-Dalsgaard2014_epsilon} use the central 3 radial modes to derive the central phase shift ($\epsilon_c$), which is more robust to acoustic glitches, whereas using the collapsed \`echelle to measure $\epsilon$ as we have done may fold in extra errors from these acoustic glitches.

We can also test the evolutionary phase using the dipole modes of these stars. Due to their mixed mode nature, the dipole oscillation modes of red giants have the important capability to probe the near core region. Successive dipole modes have a period spacing ($\Delta\Pi$) that allows us to discern whether or not the star is undergoing core burning \citep{Bedding2011_nature, Mosser2011_corot_mixed_modes}. 

For stars that have clearly identifiable dipole modes (see Section~\ref{subsec:suppression}) and do not have weakly coupled dipole modes (i.e. the dipole ridge shows mixed mode spacing), we can measure the period spacing. One can measure the period spacing by calculating the mean pairwise difference in periods of dipole modes, by aligning a period \'echelle, or using the stretched period \'echelle, which attempts to `decouple' the p-modes from the mixed-mode dipole ridge and view the g-mode contribution to them, as described in \citet{Vrard2016_pspacings,Ong2023_pspacingalgo}. The mean period spacings (often $\Delta$P$_{\rm obs}$) measured in \citet{Bedding2011_nature} and \citet{Mosser2011_corot_mixed_modes} are significantly smaller than the period spacings measured from the other two methods ($\Delta\Pi$) due to the mixed modes nearest to the pure p-mode having a different period spacing.
We measure the period spacing using the stretched period \'echelle and adopt a conservative uncertainty of 10 seconds on these estimates. 

\begin{figure}
    \centering
    \includegraphics[width=\columnwidth]{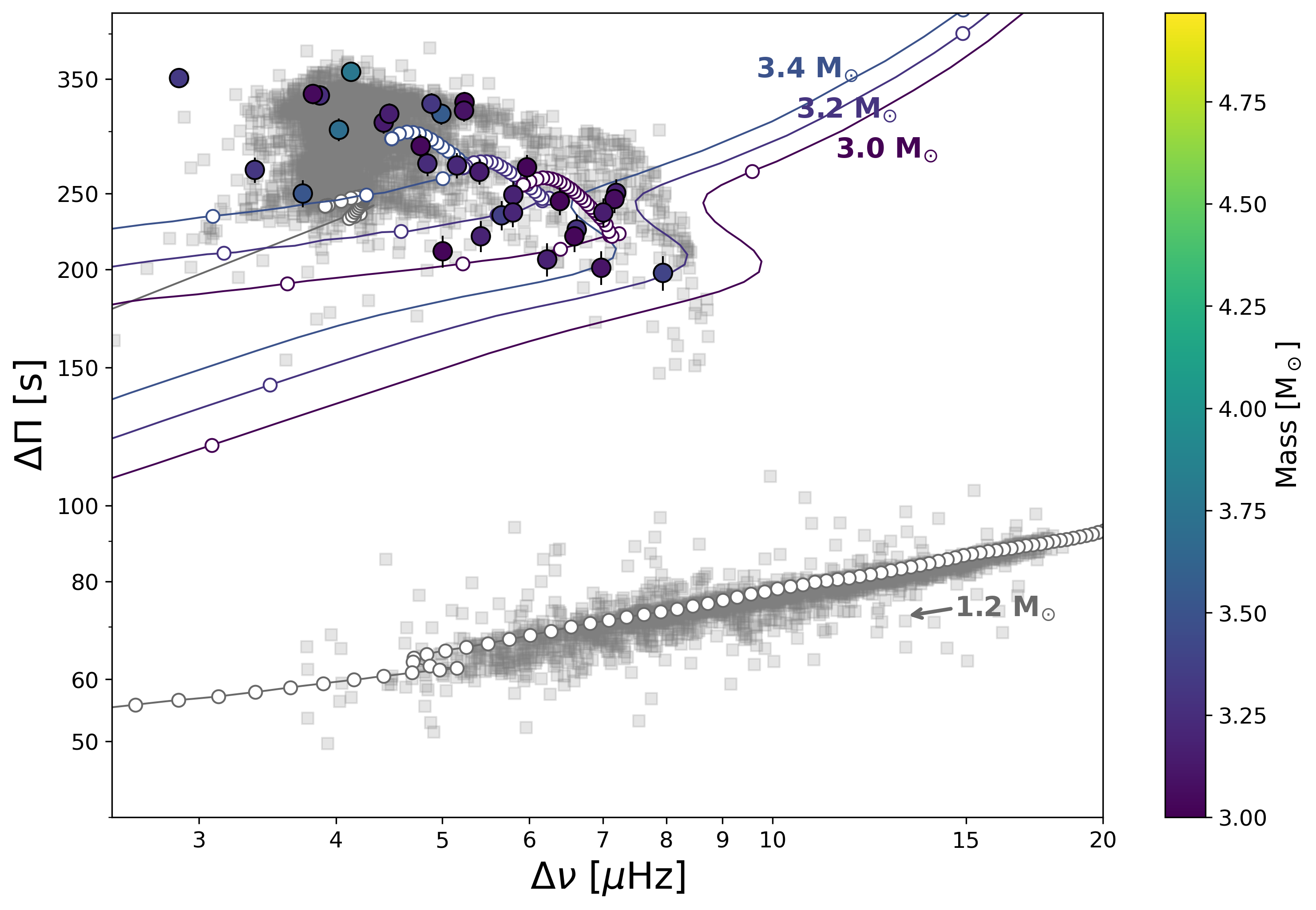}
    \caption{Here we plot the period spacing ($\Delta\Pi$) in seconds versus \Dnu{} for our stars, coloured by their masses. The error bars for $\Delta\Pi$ are standardized to 10 seconds, and the error bars for \Dnu{} are smaller than the marker size. In the background, in grey squares, we plot the values estimated for low-mass red giants in \protect\citet{Vrard2016_pspacings}. We additionally plot model evolutionary tracks of masses 1.2 \Msolar{} (grey) to represent the low-mass stars, and  3.0, 3.2, and 3.4 \Msolar{} (purple, dark blue, and light blue from the colormap) to represent the high-mass stars. These tracks are from Fig.~4 of \protect\citet{Stello2013_pspacingtracks} and points along each of them are equally spaced in time (3 million years apart). Hence, densely spaced points imply the model spends a significant amount of time in that phase.}
    \label{fig:pspacing}
\end{figure}

We compare the $\Delta\Pi$ values measured for our final sample to \citet{Vrard2016_pspacings} in Figure~\ref{fig:pspacing}. For the 20 stars which overlap in these two samples, our estimates are consistent within uncertainties (within 2\% agreement). It is clear from Figure~\ref{fig:pspacing} that our stars lie in the region occupied by typical CHeB stars. However, one must be careful when comparing the higher-mass stars of this sample to the lower-mass stars studied by \citet{Vrard2016_pspacings} or, more specifically, when comparing the secondary clump stars to the red clump stars. As shown in Fig.4 of \citet{Stello2013_pspacingtracks} and reproduced in Figure~\ref{fig:pspacing}, the predicted $\Delta\Pi$ values of high-mass models pass twice through this region of the diagram, once during the RGB phase and again in the CHeB phase. For low-mass stars, $\Delta\Pi$ is very useful for distinguishing these two evolutionary phases\footnote{We note that the 1.2 \Msolar{} track's red clump $\Delta\Pi$ estimates do not replicate the majority of the observed low-mass $\Delta\Pi$ values. This is due to the implementation of overshooting in models, as known and discussed in \citet{Constantino2015_theorydeltapi,Bossini2015_overshootpspacings}, and does not persist in the high-mass tracks.}. In the high-mass stars, $\Delta\Pi$ is less useful as the RGB and CHeB tracks are closer together. However, if the mass of the star is unambiguously known, then $\Delta\Pi$ can still be used to distinguish these two phases, because the RGB and CHeB for each track do not overlap. 
$\Delta\Pi$ serves as a probe of the near core region, which for low-mass stars is very different during the RGB and CHeB phases, because the stars have degenerate (or near-degenerate) cores on the former and not during the latter. High-mass stars, on the other hand, do not have degenerate cores at any point during their evolution, and therefore the near-core regions of RGB and CHeB are much more similar \citep{Montalban2013_pspacingconvection}.

Finally, we can approach this problem of deciding the evolutionary phase with a probabilistic approach by considering the relative lifetimes of these evolutionary phases. We use the set of evolutionary tracks provided by MIST \citep{Dotter2016_MIST0,Choi2016_MIST1} to estimate the relative amount of time each model spends in the RGB and the CHeB phases with \Dnu{} between 2 and 8 $\mu$Hz. To do this we calculate the time spent in the RGB and CHeB phases (as indicated by MIST's \texttt{phase} column) in this \Dnu{} range, and use the percentage of the CHeB phase lifetime compared to the total RGB + CHeB lifetime as a proxy for the probability that a star found here is in the CHeB phase. Note that MIST reports an estimate of \Dnu{} that it calculates as the inverse of the sound travel time across the diameter of the star, as in \citet{Christensen-Dalsgaard1988_scalingrelations, Kjeldsen1995_scalingrelations}. From this exercise, we find that solar-metallicity models with masses of 3--4 \Msolar{} spend 97--98\% of their time in the CHeB phase. For completeness, we also checked models with [Fe/H] $\pm$ 0.25 and found the same result. For comparison, a solar-metallicity 1-\Msolar{} star only spends $\sim$ 45 \% of its time as a red giant in the CHeB phase. One can see confirmation of this in Figure~\ref{fig:pspacing}, where the low-mass track has a much higher density of points in the RGB phase than do the high-mass tracks. Therefore, since the stars in our sample have masses greater than 3 \Msolar{}, it is significantly more likely for them to be CHeB secondary clump stars than RGB stars. We would only expect to approximately one star in the RGB phase for the entire sample of \totalNofstars{} stars. Therefore, our conclusion that all stars are in the CHeB phase of evolution, and belong to the secondary clump, is reasonable.


\subsection{Oscillation Envelopes}
\label{subsec:envelopes}

\begin{figure}
    \centering
    \includegraphics[width=\columnwidth]{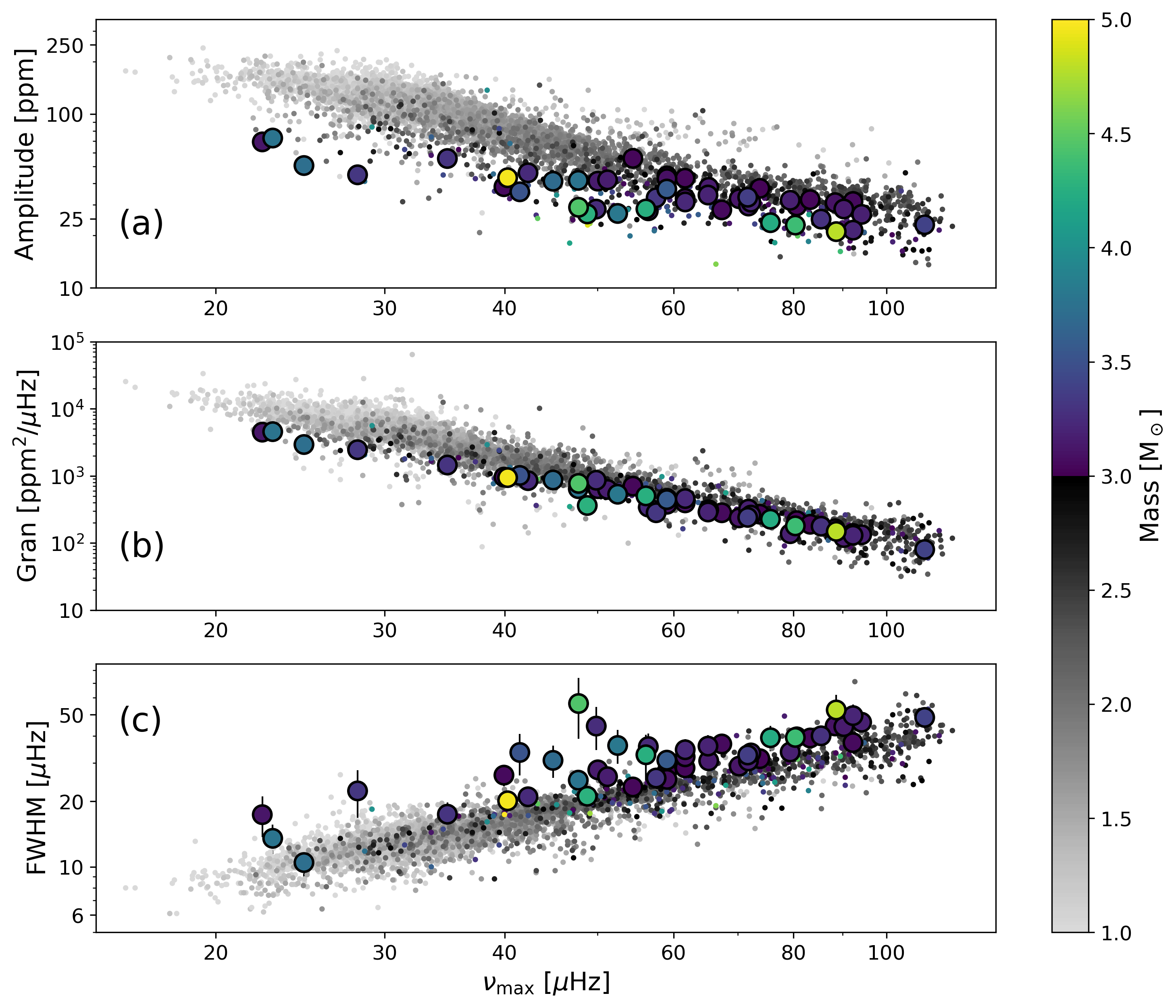}
    \caption{Three subplots describing three different parameters of the power excess all plotted against \numax{} and coloured by their masses. The upper panel shows the average mode amplitudes (see Eqn~\ref{eqn:amplitude}), the middle panel shows the granulation power at \numax{}, and the lower panel shows the FWHM of the power excess. The vertical error bars are only large enough to be visible in panel (c), showing FWHM. In the background, in small circles, we plot each of these values as estimated by \yucat{} for their red clump (CHeB) sample, coloured by their masses.}
    \label{fig:amp_width_gran}
\end{figure}

In Figure~\ref{fig:amp_width_gran} we compare three oscillation envelope parameters from our sample with the set of red clump stars studied by \yucat{}. Figure~\ref{fig:amp_width_gran}(a) shows the mean amplitude per mode for each star, defined as
\begin{equation}\label{eqn:amplitude}
    A = \frac{\sqrt{\frac{H_{\rm env} \Delta\nu}{c}}}{{{\rm sinc}\left(\frac{\pi}{2}\frac{\nu_{\rm max}}{\nu_{\rm Nyq}}\right)}},
\end{equation}
This equation includes a sinc function correction to account for the attenuation of signals approaching the Nyquist frequency due to the non-zero integration time \citep{Huber2010_keplerrgs,Murphy2012_Keplercharacteristics,Chaplin2014_supernyquist}. 
Here, H$_{\rm env}$ is the height of the oscillation excess (\texttt{A\_smooth} in the {\sc pySYD} output) and $c$ is the effective number of modes per order, adopted as 3.04 \citep{Kjeldsen2008_amplitudes}. This adopted value for $c$ is useful for comparison to the other red giant stars, but we note that many of our stars have suppressed dipole modes, and thus have fewer effective modes per order. Thus, the mean mode amplitude calculated using this method is lower for stars with suppressed dipole modes than for `normal' stars.

The trends of our mean mode amplitude estimates with \numax{} and mass are consistent with those found in \yucat{}, where the highest mass stars have much lower mode amplitudes than intermediate mass stars. 
The mean mode amplitudes estimated in this work are on average 20\% larger than those reported in \yucat{}, independent of stellar \numax{}. This is again due to the reduction in white noise in this work from adopting the PDCSAP light curves (see Section~\ref{subsec:numax}). We illustrate this using our example power density spectrum in Figure~\ref{fig:bgfit_compare}. Here we see that the reduction in the white noise component of the power density does not reduce the power in the oscillation modes, and therefore this work must estimate a larger H$_{\rm env}$, and by extension mean mode amplitude, to properly model the background and oscillations.
Our mean mode amplitudes are still consistent with the trend reported in \yucat{}, and our high mass stars have lower mean mode amplitudes than lower mass CHeB stars, as expected \citep{Mosser2012_powerexcess}.

Figure~\ref{fig:amp_width_gran}(b) shows the granulation power at \numax{}, which excludes the white noise and is also corrected for attenuation. 
These two are known to be tightly correlated, as the convective properties (and thus the granulation background) are well correlated with the acoustic cutoff frequency and therefore \numax{} \citep{Kjeldsen2011_amplitudescaling,Mathur2011_granulation,Kallinger2014_granulation}. Our results are again consistent with \yucat{}.

\begin{figure}
    \centering
    \includegraphics[width=\columnwidth]{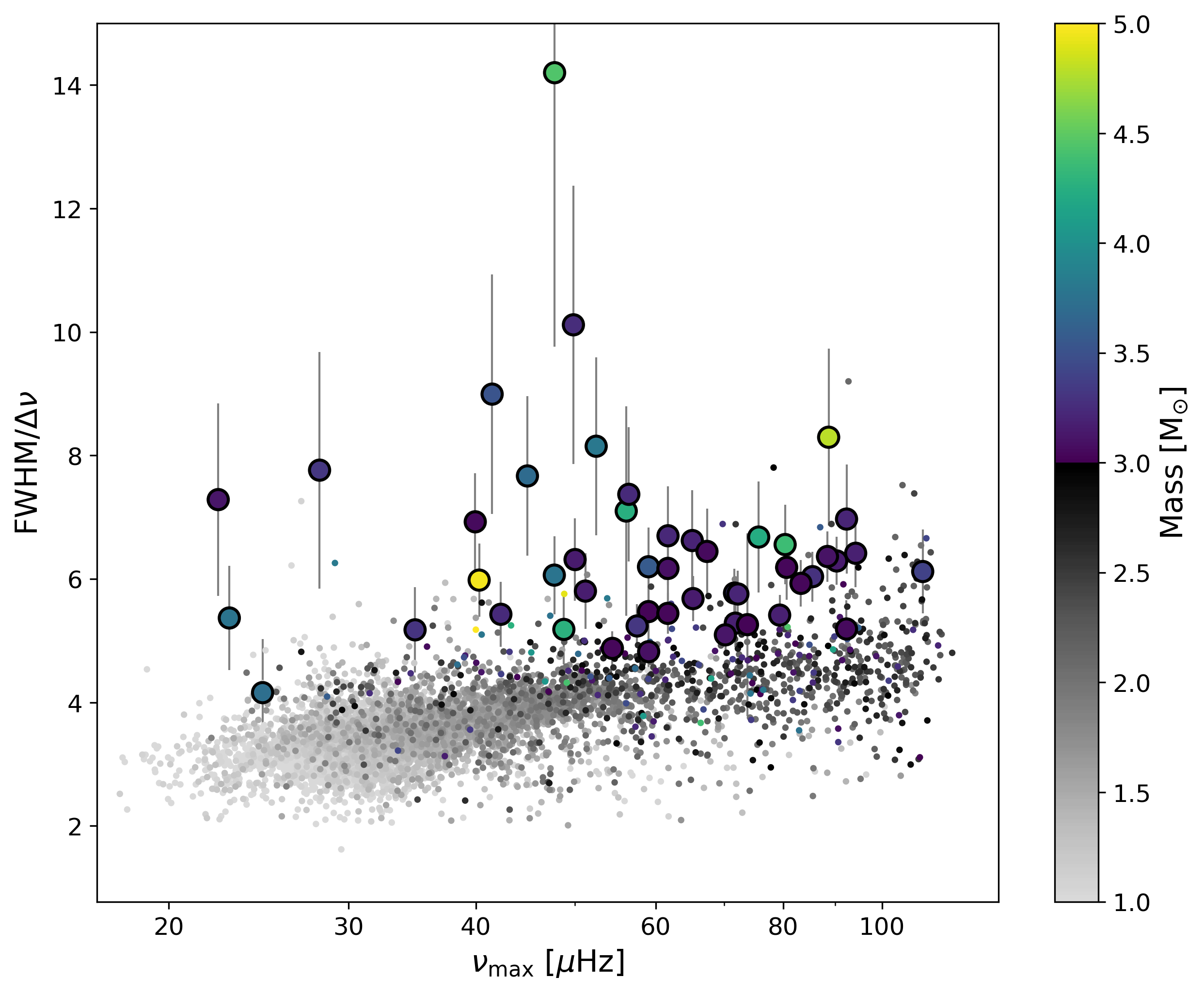}
    \caption{A plot showing the FWHM divided by \Dnu{}, which is related to an approximate number of excited orders, versus \numax{} with the markers coloured by the stellar mass. We show only the vertical error bars, for clarity. In the background, in small circles, we plot each of these values as estimated by \yucat{} for their red clump (CHeB) sample, coloured by their masses, as was done in Figure~\ref{fig:amp_width_gran}.}
    \label{fig:excited_orders}
\end{figure}

In Figure~\ref{fig:amp_width_gran}(c) we show the full-width at half-maximum (FWHM) of the power envelope, which is measured by fitting a Gaussian distribution to the power excess. The tight trend with \numax{} is also visible in Figure~\ref{fig:stacked_spectra}. Our stars have quite wide power envelopes compared to typical CHeB stars, and especially compared to the RGB stars, as previously found for high-mass stars \citep{Mosser2012_powerexcess, Yu2018_keplercatalog}. However, this effect is inflated due to the reduction in white noise once again allowing for wider envelopes, just as it allowed for larger average mode amplitudes. The error bars on the FWHM show that the few stars with anomalously large FWHM values also have relatively large uncertainties. The power excess width should increase with \numax{}, as is shown by the comparison data from \yucat{} in this figure, however that does not necessarily imply that more orders are being excited. If we divide the FWHM by \Dnu{} we can get a lower estimate of the number of excited orders, shown in Figure~\ref{fig:excited_orders}. We emphasize that this estimate is not the same nor as robust as simply counting the number of visible radial modes in the power spectrum. Our sample has on average $\sim$6 strongly excited orders using this metric, with many stars exciting at least 8 radial orders. We therefore have an excellent opportunity to study acoustic glitches with this sample (see Section~\ref{subsec:glitches}).

\subsection{Mode Visibilities}
\label{subsec:suppression}

In most red giants the radial, dipole, and quadrupole modes are easily identified. The ratios of the amplitudes for the three sets of modes, also known as their visibilities, are interesting probes of the stellar interior. One typically considers the ratios of the amplitudes of dipole modes and the quadrupole modes to those of the radial modes. Typical Kepler RGB stars have dipole mode visibilities of about $\sim$1.4 and quadrupole mode visibilities of $\sim$0.7. Some stars show strong suppression of their dipole modes, which translates to low dipole mode visibilities and has been observed to be a strong function of mass in RGB stars \citep{Mosser2012_powerexcess,Kallinger2014_granulation,Fuller2015_suppression,Cantiello2016_suppression,Stello2016_visNature,Stello2016_visPASA}.
\citet{Mosser2017_suppression_mixedmodes} studies both RGB and CHeB visibilities, but does not quote any strong differences between the two populations.

\begin{figure}
    \centering
    \includegraphics[width=\columnwidth]{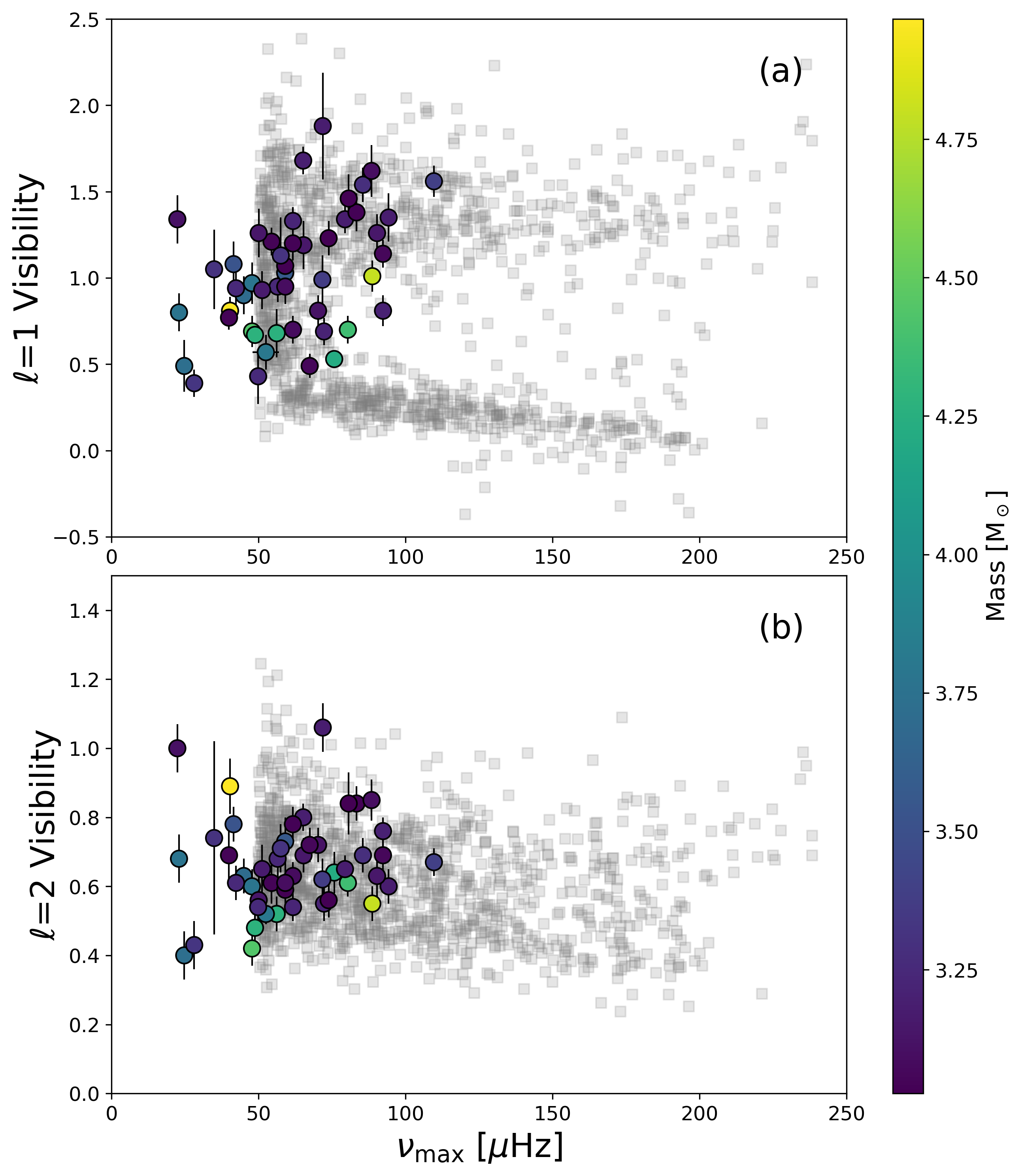}
    \caption{The visibilities for the dipole modes (panel a) and the quadrupole modes (panel b) are plotted against \numax{} for our sample, coloured by their masses. In the background, in grey squares, we plot the visibilities of red giants with masses > 1.5 \Msolar{} as estimated by \protect\citet{Stello2016_visPASA}. See discussion in Section~\ref{subsec:suppression}.}
    \label{fig:subplot_vis}
\end{figure}

We measure the dipole mode visibilities of our stars by following the framework set by \citet{Mosser2012_powerexcess} and \citet{Stello2016_visPASA,Stello2016_visNature}. However, these two groups measured the visibilities slightly differently. \citet{Mosser2012_powerexcess} measured a weighted average mode amplitude for each star by dividing a Gaussian weight, amplifying the contribution from the modes further from \numax{}. \citet{Stello2016_visPASA,Stello2016_visNature} instead used a simple, non-weighted integration of the mode amplitudes before calculating the visibility ratios. Additionally, Stello et al.\ integrated over 4 orders, whereas Mosser et al.\ integrated over 5 orders. While these two differences aren't expected to produce a systematic offset, they will nevertheless have measured slightly different values. We choose to follow the Stello et al.\ formulation, and we will be comparing to their values in our analysis. We additionally calculated the visibilities by integrating over 5 orders rather than 4 orders, and find agreement between the two cases within uncertainties. However, one critical difference from the Stello et al.\ formulation is regarding the identification of modes. For our work, we begin by measuring the frequencies of the radial modes for each star (`peak-bagging'), whereas \citet{Stello2016_visPASA} identified modes via autocorrelation with template functions at a range of \Dnu{} values. Additionally, we calculate error bars for our visibility estimates by calculating the mode visibilities for each available quarter of Kepler data and taking the standard error in the mean (standard deviation divided by the square root of the number of observed quarters). Our results are shown in Fig~\ref{fig:subplot_vis}, plotted over the estimates from \citet{Stello2016_visPASA} for comparison.

As we do not have any overlap with the sample of Stello et al.\ sample, which included only RGB stars, we tested our methodology on 10 typical stars from the Stello et al.\ sample covering a range of \numax.  We found good agreement at large \numax{} but for \numax{} in the range 50--100 $\mu$Hz, our estimates are systematically larger by about 10\%, albeit consistent within uncertainties. Whether this systematic offset is due to the Kepler data processing, the visibility calculation, or the crowdedness of the modes at low \numax{} is unclear. 

Our stars do not appear anomalous from the overall sample of red giants tested by these works. However, unlike the high-\numax{} portion of the Stello et al.\ sample, we do not see a clear distinction between our suppressed and not suppressed stars. We instead see a smooth distribution across visibility space, similar to the highest mass bin in Figure~2 of \citet{Stello2016_visNature}, and seen at the low-\numax{} end of their sample, where the suppressed theoretical model turns upwards, and therefore we do not adopt a threshold value upon which we categorise our stars as either ``suppressed" or ``not suppressed". We do confirm the trend that the highest mass stars in our sample tend to have weaker dipole mode visibilities. In fact we only measure dipole mode visibilities greater than 1.1 at masses below 3.5 \Msolar{}. Additionally, we note that there appears to be a weak correlation between \numax{} and the dipole mode visibilities. Recall that we do not see any trends with the mass of the star and its \numax{}. We believe the trend between \numax{} and dipole visibility is coincidental, especially considering that model evolutionary tracks for these high masses show that the Zero-age CHeB main sequence lies at very different values of \numax{} for each mass, and it is therefore likely that these stars are at different points in their CHeB lifetimes. We also point out that there is a weak trend in dipole mode visibilities and $\epsilon$, visible to a careful eye in Figure~\ref{fig:stacked_epsilon}. This can be interpreted as the same trend that is seen in \numax{}, as \numax{} is known to vary with \Dnu{}, and we have shown that \Dnu{} also correlates with $\epsilon$ in Figure~\ref{fig:epsilon}. Thus we are skeptical that this trend is more than a small number statistical effect.

For quadrupole modes, suppression is a much more subtle effect. As shown in \citet{Stello2016_visPASA}, the quadrupole mode suppression is much more clear at large values of \numax{}. This is also clear in the lower panel of Figure~\ref{fig:subplot_vis} where the estimates from \citet{Stello2016_visPASA} are plotted. At low \numax{} there is very little bimodality in the quadrupole mode visibility. Therefore it is not necessarily surprising that we see a very localized distribution of the $\ell$=2 visibilities for our sample. However, we do generally see that the highest-mass stars in our sample tend to have quadrupole visibilities on the low end of our distribution.

Of the non-radial modes, the dipole modes are more strongly coupled to the core \citep{Dupret2009_modepredictions}, and therefore probe the interior more strongly, and may be more affected by core characteristics, which are the main difference between the RGB and the CHeB stars. As discussed in Section~\ref{subsec:evol_phase}, our sample is expected to be dominated by CHeB stars, whose visibilities have been discussed theoretically by \citet{Cantiello2016_suppression} and observed by \citet{Mosser2017_suppression_mixedmodes}.
The leading theory is that the suppression of these non-radial modes in RGB stars is due to fossil magnetic fields left from the convective cores of the main sequence phase \citep{Fuller2015_suppression,Stello2016_visNature,Stello2016_visPASA,Cantiello2016_suppression}. This ``magnetic greenhouse effect" \citep{Fuller2015_suppression} arises as the mode energy that leaks into the g-mode cavity is dissipated due to the core magnetic field. \citet{Cantiello2016_suppression} describes three possible scenarios for suppression in CHeB stars, relating to whether or not the fossil magnetic field is destroyed in the helium flashes of intermediate-mass stars. Our stars will not exhibit helium flashes before the CHeB phase, and thus the core magnetic field should not be affected. 
Additionally, our sample of high-mass CHeB stars should have convective cores during the current phase, so perhaps they are regenerating a magnetic dynamo in their core, driving further suppression. However, there are other recent works such as \citet{Mosser2017_suppression_mixedmodes} which dispute the magnetic greenhouse effect, showing that suppression does not disrupt mixed modes as one would expect from the magnetic greenhouse. The effects of magnetic fields on the g-mode cavities in the cores of red giants have begun to be explored by works such as \citet{Loi2018_magneticsuppression_theory}, and our high-mass sample provides an interesting lens through which to study the source of dipole mode suppression.

\subsection{Radial Mode Glitches}
\label{subsec:glitches}

Deviations of \Dnu{} from the asymptotic relation are often called acoustic ``glitches", and are indicative of abrupt structural changes within the envelope of the star, such as the edges of partial ionization zones (especially the Helium partial ionization zone) and the bottom of the convection zone \citep{Vrard2015_glitches}. Perhaps owing to their wide excitation envelopes and thus high number of excited orders (see Section~\ref{subsec:envelopes} and Figure~\ref{fig:amp_width_gran}), all stars in our sample show radial mode glitches. Additionally, many of our stars share a similar glitch morphology, which may point to shared structure within the secondary clump. In this work we do not measure properties (amplitude, phase, or period) of radial mode glitches. However, we note that we can see the glitch effect in the broadening of the collapsed radial mode ridge in Figure~\ref{fig:stacked_epsilon}, for example. This also adds an extra uncertainty onto the measured epsilon value (see Figure~\ref{fig:epsilon}). We are also sensitive to glitches in our visibility calculation, but we account for these by peak-bagging the radial modes, rather than assuming they are spaced exactly \Dnu{} from each other. We intend to further study these glitches in later work.

\section{Summary}
\label{sec:conclusions}

In this work we present a sample of Kepler red giant stars originating from the full sample of 16,000 stars published in \yucat{}, focusing on the highest mass stars within the field. This sample of \Nofstars{} stars all have estimated masses greater than 3 \Msolar{}, as calculated by the asteroseismic scaling relations (Eqn~\ref{eqn:mass_scaling}). In creating this sample, we have calculated improved oscillation spectra using Kepler DR25 PDCSAP light curves and improved background fits using the {\sc pySYD} program \citep{Chontos2022_pysyd}. We additionally pay special attention to which \Teff{} is adopted for the sample, as that is an important source of systematic uncertainty within the mass estimates. We adopted APOGEE \citep{APOGEE_DR16} or LAMOST \citep{LAMOST} temperatures when available, and those from \yucat{} otherwise. In future works we intend to further discuss the spectroscopic fits for these stars. Our mass uncertainties are also affected by the uncertainties on the global oscillation parameters \numax{} and \Dnu{}, which are hard to avoid due to the stochastic nature of red giant oscillations.
We have also provided discussion on the likelihood of stellar contamination in our data, and we do not expect this to be an issue for our sample.

The vast majority of our stars are near the lower mass cutoff of 3 \Msolar{}, with $\sim$ 58 \% of our stars having masses of 3--3.25 \Msolar{}, but we do find 7 stars with mass estimates greater than 4 \Msolar{}. These stars in particular are excellent probes to study the seismic parameters of high-mass red giants, and we will focus on them in particular in future works. 
The parameters of the power excess envelope for our stars are consistent with other CHeB stars from \yucat{} when accounting for the differences in background fitting and power spectra between these two samples. The stars presented here have a large number of excited radial orders due to their wide envelopes, and thus are unique laboratories for studying radial mode glitches and the internal ionization zones of CHeB stars. We leave the study of these glitches to future works.

One of the most important details of a high-mass star is its evolutionary phase, as this governs the interpretation of most oscillation parameters. The general term ``red giants" encompasses the H-shell burning RGB stars, the CHeB red clump (secondary clump at these masses) stars, and the He-shell burning AGB stars. Based on the relative lifetimes of high-mass MIST models \citep{Dotter2016_MIST0,Choi2016_MIST1} in the RGB and CHeB phases, we do not expect more than one star in our sample to be in the RGB phase. There are a few seismic parameters we can use as probes into evolutionary phase to check this assumption, however. We investigate the phase shift of the radial modes, $\epsilon$, as a function of \Dnu{}, which is known to be different for these two evolutionary phases due to differences in the surface layer of the star \citep{Bedding2011_nature,Kallinger2012_epsilon,Christensen-Dalsgaard2014_epsilon}. We also investigate the mixed mode period spacings, $\Delta\Pi$, when possible, however we caution that this value is known to not show as strong of differences at high mass, since the core region of high-mass RGBs is not degenerate \citep{Stello2013_pspacingtracks}. Nonetheless, our stars show $\Delta\Pi$ values consistent with CHeB stars.

The relative visibility of the dipole and quadrupole modes to the radial modes is also a quite interesting parameter to investigate. The dipole mode visibility in particular is known to be a strong function of mass in RGB stars \citep{Mosser2012_powerexcess,Fuller2015_suppression,Stello2016_visNature,Stello2016_visPASA,Cantiello2016_suppression}, with higher mass RGB stars tending to show more strongly suppressed dipole modes than their lower mass brethren. In our sample, all stars with masses greater than 3.5 \Msolar{} show dipole mode visibilities weaker than 1.1, which is consistent with the findings from RGB stars. We do not see a strong separation between the suppressed dipole modes and the strong dipole modes, instead seeing a roughly smooth number distribution in visibility space. We see no anomalies in the quadrupole mode visibilities. If the dipole mode suppression seen in red giants is due to fossil magnetic fields, as suggested by \citet{Fuller2015_suppression}, then these high-mass CHeB stars provide a unique probe, as they may have convective cores which may excite their own magnetic dynamos.

This paper is Part I of our series of works studying these high-mass Kepler red giants. In the future we intend to dive into the spectroscopic details of these stars as well as provide detailed asteroseismic models of a few of these stars. These will both provide strong anchor points for this mass regime which has been traditionally avoided. These may also provide independent anchor points useful for the study of their progenitor main sequence stars such as the slowly-pulsating B stars and $\delta$ Scuti stars, as well as their final stages, the white dwarf stars (see \citealt{Kurtz2022_review} for review). Modeling of these highest mass stars will also allow us to test the scaling relations at extreme mass, which will provide constraints on the usefulness of these scaling relations.


\section*{Acknowledgements}

CC would like to thank the attendees of TASC7/KASC14 for their helpful comments on the poster regarding this work, as well as Beno\^it Mosser and Karsten Brogaard for comments during the drafting stage. 
We thank the anonymous referees for their comments.
We gratefully acknowledge support from the Australian Research Council through Discovery Projects DP190100666 and DP210103119 and Laureate Fellowship FL220100117.
This paper includes data collected by the Kepler mission and obtained from the MAST data archive at the Space Telescope Science Institute (STScI). Funding for the Kepler mission is provided by the NASA Science Mission Directorate. STScI is operated by the Association of Universities for Research in Astronomy, Inc., under NASA contract NAS 5–26555.

MIST is built using Modules for Experiments in Stellar Astrophysics
\citep[MESA][]{Paxton2011, Paxton2013, Paxton2015, Paxton2018, Paxton2019, Jermyn2023}. The MESA EOS is a blend of the OPAL \citep{Rogers2002}, SCVH
\citep{Saumon1995}, FreeEOS \citep{Irwin2004}, HELM \citep{Timmes2000},
PC \citep{Potekhin2010}, and Skye \citep{Jermyn2021} EOSes.
Radiative opacities are primarily from OPAL \citep{Iglesias1993,
Iglesias1996}, with low-temperature data from \citet{Ferguson2005}
and the high-temperature, Compton-scattering dominated regime by
\citet{Poutanen2017}.  Electron conduction opacities are from
\citet{Cassisi2007} and \citet{Blouin2020}.
Nuclear reaction rates are from JINA REACLIB \citep{Cyburt2010}, NACRE \citep{Angulo1999} and
additional tabulated weak reaction rates \citet{Fuller1985, Oda1994,
Langanke2000}.  Screening is included via the prescription of \citet{Chugunov2007}.
Thermal neutrino loss rates are from \citet{Itoh1996}.
This work made use of several publicly available {\tt python} packages: {\tt astropy} \citep{astropy:2013,astropy:2018}, 
{\tt lightkurve} \citep{lightkurve2018},
{\tt matplotlib} \citep{matplotlib2007}, 
{\tt numpy} \citep{numpy2020}, and 
{\tt scipy} \citep{scipy2020}.

\section*{Data Availability}

The full version of Table 1 can be accessed online at [future online upload link]. The light curves and power spectra generated for this work can be requested from the corresponding author.

\typeout{}
\bibliographystyle{mnras}
\bibliography{highmass_rc,mesa}

\begin{thebibliography}{}
\makeatletter
\relax
\def\mn@urlcharsother{\let\do\@makeother \do\$\do\&\do\#\do\^\do\_\do\%\do\~}
\def\mn@doi{\begingroup\mn@urlcharsother \@ifnextchar [ {\mn@doi@}
  {\mn@doi@[]}}
\def\mn@doi@[#1]#2{\def\@tempa{#1}\ifx\@tempa\@empty \href
  {http://dx.doi.org/#2} {doi:#2}\else \href {http://dx.doi.org/#2} {#1}\fi
  \endgroup}
\def\mn@eprint#1#2{\mn@eprint@#1:#2::\@nil}
\def\mn@eprint@arXiv#1{\href {http://arxiv.org/abs/#1} {{\tt arXiv:#1}}}
\def\mn@eprint@dblp#1{\href {http://dblp.uni-trier.de/rec/bibtex/#1.xml}
  {dblp:#1}}
\def\mn@eprint@#1:#2:#3:#4\@nil{\def\@tempa {#1}\def\@tempb {#2}\def\@tempc
  {#3}\ifx \@tempc \@empty \let \@tempc \@tempb \let \@tempb \@tempa \fi \ifx
  \@tempb \@empty \def\@tempb {arXiv}\fi \@ifundefined
  {mn@eprint@\@tempb}{\@tempb:\@tempc}{\expandafter \expandafter \csname
  mn@eprint@\@tempb\endcsname \expandafter{\@tempc}}}

\bibitem[\protect\citeauthoryear{{Angulo} et~al.,}{{Angulo}
  et~al.}{1999}]{Angulo1999}
{Angulo} C.,  et~al., 1999, \mn@doi [\nphysa] {10.1016/S0375-9474(99)00030-5},
  \href {https://ui.adsabs.harvard.edu/abs/1999NuPhA.656....3A} {656, 3}

\bibitem[\protect\citeauthoryear{{Astropy Collaboration}}{{Astropy
  Collaboration}}{2013}]{astropy:2013}
{Astropy Collaboration} 2013, \mn@doi [\aap] {10.1051/0004-6361/201322068},
  \href {http://adsabs.harvard.edu/abs/2013A%26A...558A..33A} {558, A33}

\bibitem[\protect\citeauthoryear{{Astropy Collaboration}}{{Astropy
  Collaboration}}{2018}]{astropy:2018}
{Astropy Collaboration} 2018, \mn@doi [\aj] {10.3847/1538-3881/aabc4f}, \href
  {https://ui.adsabs.harvard.edu/abs/2018AJ....156..123A} {156, 123}

\bibitem[\protect\citeauthoryear{{Bedding} \& {Kjeldsen}}{{Bedding} \&
  {Kjeldsen}}{2003}]{Bedding+Kjeldsen2003_review}
{Bedding} T.~R.,  {Kjeldsen} H.,  2003, \mn@doi [\pasa] {10.1071/AS03025},
  \href {https://ui.adsabs.harvard.edu/abs/2003PASA...20..203B} {20, 203}

\bibitem[\protect\citeauthoryear{{Bedding} et~al.,}{{Bedding}
  et~al.}{2010}]{Bedding2010_firstkepler}
{Bedding} T.~R.,  et~al., 2010, \mn@doi [\apjl] {10.1088/2041-8205/713/2/L176},
  \href {https://ui.adsabs.harvard.edu/abs/2010ApJ...713L.176B} {713, L176}

\bibitem[\protect\citeauthoryear{{Bedding} et~al.,}{{Bedding}
  et~al.}{2011}]{Bedding2011_nature}
{Bedding} T.~R.,  et~al., 2011, \mn@doi [\nat] {10.1038/nature09935}, \href
  {https://ui.adsabs.harvard.edu/abs/2011Natur.471..608B} {471, 608}

\bibitem[\protect\citeauthoryear{{Blouin}, {Shaffer}, {Saumon}  \&
  {Starrett}}{{Blouin} et~al.}{2020}]{Blouin2020}
{Blouin} S.,  {Shaffer} N.~R.,  {Saumon} D.,   {Starrett} C.~E.,  2020, \mn@doi
  [\apj] {10.3847/1538-4357/ab9e75}, \href
  {https://ui.adsabs.harvard.edu/abs/2020ApJ...899...46B} {899, 46}

\bibitem[\protect\citeauthoryear{{Bossini} et~al.,}{{Bossini}
  et~al.}{2015}]{Bossini2015_overshootpspacings}
{Bossini} D.,  et~al., 2015, \mn@doi [\mnras] {10.1093/mnras/stv1738}, \href
  {https://ui.adsabs.harvard.edu/abs/2015MNRAS.453.2290B} {453, 2290}

\bibitem[\protect\citeauthoryear{{Brogaard} et~al.,}{{Brogaard}
  et~al.}{2016}]{Brogaard2016_testingscaling}
{Brogaard} K.,  et~al., 2016, \mn@doi [Astronomische Nachrichten]
  {10.1002/asna.201612374}, \href
  {https://ui.adsabs.harvard.edu/abs/2016AN....337..793B} {337, 793}

\bibitem[\protect\citeauthoryear{{Brogaard} et~al.,}{{Brogaard}
  et~al.}{2022}]{Brogaard2022_testingscaling}
{Brogaard} K.,  et~al., 2022, \mn@doi [\aap] {10.1051/0004-6361/202244345},
  \href {https://ui.adsabs.harvard.edu/abs/2022A&A...668A..82B} {668, A82}

\bibitem[\protect\citeauthoryear{{Brown}, {Gilliland}, {Noyes}  \&
  {Ramsey}}{{Brown} et~al.}{1991}]{Brown1991_numaxscaling}
{Brown} T.~M.,  {Gilliland} R.~L.,  {Noyes} R.~W.,   {Ramsey} L.~W.,  1991,
  \mn@doi [\apj] {10.1086/169725}, \href
  {https://ui.adsabs.harvard.edu/abs/1991ApJ...368..599B} {368, 599}

\bibitem[\protect\citeauthoryear{{Campante} et~al.,}{{Campante}
  et~al.}{2017}]{Campante2017_PSPS}
{Campante} T.~L.,  et~al., 2017, \mn@doi [\mnras] {10.1093/mnras/stx876}, \href
  {https://ui.adsabs.harvard.edu/abs/2017MNRAS.469.1360C} {469, 1360}

\bibitem[\protect\citeauthoryear{{Cantiello}, {Fuller}  \&
  {Bildsten}}{{Cantiello} et~al.}{2016}]{Cantiello2016_suppression}
{Cantiello} M.,  {Fuller} J.,   {Bildsten} L.,  2016, \mn@doi [\apj]
  {10.3847/0004-637X/824/1/14}, \href
  {https://ui.adsabs.harvard.edu/abs/2016ApJ...824...14C} {824, 14}

\bibitem[\protect\citeauthoryear{{Cassisi}, {Potekhin}, {Pietrinferni},
  {Catelan}  \& {Salaris}}{{Cassisi} et~al.}{2007}]{Cassisi2007}
{Cassisi} S.,  {Potekhin} A.~Y.,  {Pietrinferni} A.,  {Catelan} M.,   {Salaris}
  M.,  2007, \mn@doi [\apj] {10.1086/516819}, \href
  {https://ui.adsabs.harvard.edu/abs/2007ApJ...661.1094C} {661, 1094}

\bibitem[\protect\citeauthoryear{{Chaplin} \& {Miglio}}{{Chaplin} \&
  {Miglio}}{2013}]{Chaplin+Miglio2013_review}
{Chaplin} W.~J.,  {Miglio} A.,  2013, \mn@doi [\araa]
  {10.1146/annurev-astro-082812-140938}, \href
  {https://ui.adsabs.harvard.edu/abs/2013ARA&A..51..353C} {51, 353}

\bibitem[\protect\citeauthoryear{{Chaplin}, {Elsworth}, {Davies}, {Campante},
  {Handberg}, {Miglio}  \& {Basu}}{{Chaplin}
  et~al.}{2014}]{Chaplin2014_supernyquist}
{Chaplin} W.~J.,  {Elsworth} Y.,  {Davies} G.~R.,  {Campante} T.~L.,
  {Handberg} R.,  {Miglio} A.,   {Basu} S.,  2014, \mn@doi [\mnras]
  {10.1093/mnras/stu1811}, \href
  {https://ui.adsabs.harvard.edu/abs/2014MNRAS.445..946C} {445, 946}

\bibitem[\protect\citeauthoryear{{Choi}, {Dotter}, {Conroy}, {Cantiello},
  {Paxton}  \& {Johnson}}{{Choi} et~al.}{2016}]{Choi2016_MIST1}
{Choi} J.,  {Dotter} A.,  {Conroy} C.,  {Cantiello} M.,  {Paxton} B.,
  {Johnson} B.~D.,  2016, \mn@doi [\apj] {10.3847/0004-637X/823/2/102}, \href
  {https://ui.adsabs.harvard.edu/abs/2016ApJ...823..102C} {823, 102}

\bibitem[\protect\citeauthoryear{{Chontos}, {Huber}, {Sayeed}  \&
  {Yamsiri}}{{Chontos} et~al.}{2022}]{Chontos2022_pysyd}
{Chontos} A.,  {Huber} D.,  {Sayeed} M.,   {Yamsiri} P.,  2022, \mn@doi [The
  Journal of Open Source Software] {10.21105/joss.03331}, \href
  {https://ui.adsabs.harvard.edu/abs/2022JOSS....7.3331C} {7, 3331}

\bibitem[\protect\citeauthoryear{{Christensen-Dalsgaard}}{{Christensen-Dalsgaard}}{1988}]{Christensen-Dalsgaard1988_scalingrelations}
{Christensen-Dalsgaard} J.,  1988, in {Christensen-Dalsgaard} J.,  {Frandsen}
  S.,  eds,  IAU Symposium Vol. 123, Advances in Helio- and Asteroseismology.
  p.~3

\bibitem[\protect\citeauthoryear{{Christensen-Dalsgaard}, {Silva Aguirre},
  {Elsworth}  \& {Hekker}}{{Christensen-Dalsgaard}
  et~al.}{2014}]{Christensen-Dalsgaard2014_epsilon}
{Christensen-Dalsgaard} J.,  {Silva Aguirre} V.,  {Elsworth} Y.,   {Hekker} S.,
   2014, \mn@doi [\mnras] {10.1093/mnras/stu2007}, \href
  {https://ui.adsabs.harvard.edu/abs/2014MNRAS.445.3685C} {445, 3685}

\bibitem[\protect\citeauthoryear{{Chugunov}, {Dewitt}  \&
  {Yakovlev}}{{Chugunov} et~al.}{2007}]{Chugunov2007}
{Chugunov} A.~I.,  {Dewitt} H.~E.,   {Yakovlev} D.~G.,  2007, \mn@doi [\prd]
  {10.1103/PhysRevD.76.025028}, \href
  {https://ui.adsabs.harvard.edu/abs/2007PhRvD..76b5028C} {76, 025028}

\bibitem[\protect\citeauthoryear{{Colman} et~al.,}{{Colman}
  et~al.}{2017}]{Colman2017_anamolouspeaks}
{Colman} I.~L.,  et~al., 2017, \mn@doi [\mnras] {10.1093/mnras/stx1056}, \href
  {https://ui.adsabs.harvard.edu/abs/2017MNRAS.469.3802C} {469, 3802}

\bibitem[\protect\citeauthoryear{{Constantino}, {Campbell},
  {Christensen-Dalsgaard}, {Lattanzio}  \& {Stello}}{{Constantino}
  et~al.}{2015}]{Constantino2015_theorydeltapi}
{Constantino} T.,  {Campbell} S.~W.,  {Christensen-Dalsgaard} J.,  {Lattanzio}
  J.~C.,   {Stello} D.,  2015, \mn@doi [\mnras] {10.1093/mnras/stv1264}, \href
  {https://ui.adsabs.harvard.edu/abs/2015MNRAS.452..123C} {452, 123}

\bibitem[\protect\citeauthoryear{{Corsaro} et~al.,}{{Corsaro}
  et~al.}{2012}]{Corsaro2012_epsilonrelation}
{Corsaro} E.,  et~al., 2012, \mn@doi [\apj] {10.1088/0004-637X/757/2/190},
  \href {https://ui.adsabs.harvard.edu/abs/2012ApJ...757..190C} {757, 190}

\bibitem[\protect\citeauthoryear{{Cui} et~al.,}{{Cui} et~al.}{2012}]{LAMOST}
{Cui} X.-Q.,  et~al., 2012, \mn@doi [Research in Astronomy and Astrophysics]
  {10.1088/1674-4527/12/9/003}, \href
  {https://ui.adsabs.harvard.edu/abs/2012RAA....12.1197C} {12, 1197}

\bibitem[\protect\citeauthoryear{{Cyburt} et~al.,}{{Cyburt}
  et~al.}{2010}]{Cyburt2010}
{Cyburt} R.~H.,  et~al., 2010, \mn@doi [\apjs] {10.1088/0067-0049/189/1/240},
  \href {https://ui.adsabs.harvard.edu/abs/2010ApJS..189..240C} {189, 240}

\bibitem[\protect\citeauthoryear{{Dhanpal}, {Benomar}, {Hanasoge}, {Kundu},
  {Dhuri}, {Das}  \& {Kaul}}{{Dhanpal}
  et~al.}{2022}]{Dhanpal2022_MLseparations}
{Dhanpal} S.,  {Benomar} O.,  {Hanasoge} S.,  {Kundu} A.,  {Dhuri} D.,  {Das}
  D.,   {Kaul} B.,  2022, \mn@doi [\apj] {10.3847/1538-4357/ac5247}, \href
  {https://ui.adsabs.harvard.edu/abs/2022ApJ...928..188D} {928, 188}

\bibitem[\protect\citeauthoryear{{Dotter}}{{Dotter}}{2016}]{Dotter2016_MIST0}
{Dotter} A.,  2016, \mn@doi [\apjs] {10.3847/0067-0049/222/1/8}, \href
  {https://ui.adsabs.harvard.edu/abs/2016ApJS..222....8D} {222, 8}

\bibitem[\protect\citeauthoryear{{Dupret} et~al.,}{{Dupret}
  et~al.}{2009}]{Dupret2009_modepredictions}
{Dupret} M.~A.,  et~al., 2009, \mn@doi [\aap] {10.1051/0004-6361/200911713},
  \href {https://ui.adsabs.harvard.edu/abs/2009A&A...506...57D} {506, 57}

\bibitem[\protect\citeauthoryear{{Fabricius} et~al.,}{{Fabricius}
  et~al.}{2021}]{Fabricius2021_gaiaedr3validation}
{Fabricius} C.,  et~al., 2021, \mn@doi [\aap] {10.1051/0004-6361/202039834},
  \href {https://ui.adsabs.harvard.edu/abs/2021A&A...649A...5F} {649, A5}

\bibitem[\protect\citeauthoryear{{Ferguson}, {Alexander}, {Allard}, {Barman},
  {Bodnarik}, {Hauschildt}, {Heffner-Wong}  \& {Tamanai}}{{Ferguson}
  et~al.}{2005}]{Ferguson2005}
{Ferguson} J.~W.,  {Alexander} D.~R.,  {Allard} F.,  {Barman} T.,  {Bodnarik}
  J.~G.,  {Hauschildt} P.~H.,  {Heffner-Wong} A.,   {Tamanai} A.,  2005,
  \mn@doi [\apj] {10.1086/428642}, \href
  {https://ui.adsabs.harvard.edu/abs/2005ApJ...623..585F} {623, 585}

\bibitem[\protect\citeauthoryear{{Fuller}, {Fowler}  \& {Newman}}{{Fuller}
  et~al.}{1985}]{Fuller1985}
{Fuller} G.~M.,  {Fowler} W.~A.,   {Newman} M.~J.,  1985, \mn@doi [\apj]
  {10.1086/163208}, \href
  {https://ui.adsabs.harvard.edu/abs/1985ApJ...293....1F} {293, 1}

\bibitem[\protect\citeauthoryear{{Fuller}, {Cantiello}, {Stello}, {Garcia}  \&
  {Bildsten}}{{Fuller} et~al.}{2015}]{Fuller2015_suppression}
{Fuller} J.,  {Cantiello} M.,  {Stello} D.,  {Garcia} R.~A.,   {Bildsten} L.,
  2015, \mn@doi [Science] {10.1126/science.aac6933}, \href
  {https://ui.adsabs.harvard.edu/abs/2015Sci...350..423F} {350, 423}

\bibitem[\protect\citeauthoryear{{Gaia Collaboration}}{{Gaia
  Collaboration}}{2022}]{GaiaDR3_nonsinglestarcatalog}
{Gaia Collaboration} 2022, VizieR Online Data Catalog, \href
  {https://ui.adsabs.harvard.edu/abs/2022yCat.1357....0G} {p. I/357}

\bibitem[\protect\citeauthoryear{{Gaia Collaboration} et~al.,}{{Gaia
  Collaboration} et~al.}{2018}]{Gaia_DR2}
{Gaia Collaboration} et~al., 2018, \mn@doi [\aap]
  {10.1051/0004-6361/201833051}, \href
  {https://ui.adsabs.harvard.edu/abs/2018A&A...616A...1G} {616, A1}

\bibitem[\protect\citeauthoryear{{Gaia Collaboration} et~al.,}{{Gaia
  Collaboration} et~al.}{2023}]{GaiaDR3_release}
{Gaia Collaboration} et~al., 2023, \mn@doi [\aap]
  {10.1051/0004-6361/202243940}, \href
  {https://ui.adsabs.harvard.edu/abs/2023A&A...674A...1G} {674, A1}

\bibitem[\protect\citeauthoryear{{Garc{\'\i}a} \& {Ballot}}{{Garc{\'\i}a} \&
  {Ballot}}{2019}]{Garcia2019_review_solar_like}
{Garc{\'\i}a} R.~A.,  {Ballot} J.,  2019, \mn@doi [Living Reviews in Solar
  Physics] {10.1007/s41116-019-0020-1}, \href
  {https://ui.adsabs.harvard.edu/abs/2019LRSP...16....4G} {16, 4}

\bibitem[\protect\citeauthoryear{{Gaulme}, {Jackiewicz}, {Appourchaux}  \&
  {Mosser}}{{Gaulme} et~al.}{2014}]{Gaulme2014_binarysuppression}
{Gaulme} P.,  {Jackiewicz} J.,  {Appourchaux} T.,   {Mosser} B.,  2014, \mn@doi
  [\apj] {10.1088/0004-637X/785/1/5}, \href
  {https://ui.adsabs.harvard.edu/abs/2014ApJ...785....5G} {785, 5}

\bibitem[\protect\citeauthoryear{{Gaulme} et~al.,}{{Gaulme}
  et~al.}{2016}]{Gaulme2016_testingscaling}
{Gaulme} P.,  et~al., 2016, \mn@doi [\apj] {10.3847/0004-637X/832/2/121}, \href
  {https://ui.adsabs.harvard.edu/abs/2016ApJ...832..121G} {832, 121}

\bibitem[\protect\citeauthoryear{{Girardi}}{{Girardi}}{1999}]{Girardi1999_secondaryclump}
{Girardi} L.,  1999, \mn@doi [\mnras] {10.1046/j.1365-8711.1999.02746.x}, \href
  {https://ui.adsabs.harvard.edu/abs/1999MNRAS.308..818G} {308, 818}

\bibitem[\protect\citeauthoryear{{Handberg}, {Brogaard}, {Miglio}, {Bossini},
  {Elsworth}, {Slumstrup}, {Davies}  \& {Chaplin}}{{Handberg}
  et~al.}{2017}]{Handberg2017_masslosscluster}
{Handberg} R.,  {Brogaard} K.,  {Miglio} A.,  {Bossini} D.,  {Elsworth} Y.,
  {Slumstrup} D.,  {Davies} G.~R.,   {Chaplin} W.~J.,  2017, \mn@doi [\mnras]
  {10.1093/mnras/stx1929}, \href
  {https://ui.adsabs.harvard.edu/abs/2017MNRAS.472..979H} {472, 979}

\bibitem[\protect\citeauthoryear{Harris et~al.,}{Harris
  et~al.}{2020}]{numpy2020}
Harris C.~R.,  et~al., 2020, \mn@doi [Nature] {10.1038/s41586-020-2649-2}, 585,
  357

\bibitem[\protect\citeauthoryear{{Harvey}}{{Harvey}}{1985}]{Harvey1985}
{Harvey} J.,  1985, in {Rolfe} E.,  {Battrick} B.,  eds,  ESA Special
  Publication Vol. 235, Future Missions in Solar, Heliospheric \& Space Plasma
  Physics. p.~199

\bibitem[\protect\citeauthoryear{{Hekker} \& {Christensen-Dalsgaard}}{{Hekker}
  \& {Christensen-Dalsgaard}}{2017}]{Hekker2017_review}
{Hekker} S.,  {Christensen-Dalsgaard} J.,  2017, \mn@doi [\aapr]
  {10.1007/s00159-017-0101-x}, \href
  {https://ui.adsabs.harvard.edu/abs/2017A&ARv..25....1H} {25, 1}

\bibitem[\protect\citeauthoryear{{Hekker} et~al.,}{{Hekker}
  et~al.}{2011}]{Hekker2011_comparison}
{Hekker} S.,  et~al., 2011, \mn@doi [\aap] {10.1051/0004-6361/201015185}, \href
  {https://ui.adsabs.harvard.edu/abs/2011A&A...525A.131H} {525, A131}

\bibitem[\protect\citeauthoryear{{Hon} et~al.,}{{Hon}
  et~al.}{2021}]{Hon2021_tess_qlp}
{Hon} M.,  et~al., 2021, \mn@doi [\apj] {10.3847/1538-4357/ac14b1}, \href
  {https://ui.adsabs.harvard.edu/abs/2021ApJ...919..131H} {919, 131}

\bibitem[\protect\citeauthoryear{{Howell}, {Campbell}, {Stello}  \& {De
  Silva}}{{Howell} et~al.}{2022}]{Howell2022_masslossM4}
{Howell} M.,  {Campbell} S.~W.,  {Stello} D.,   {De Silva} G.~M.,  2022,
  \mn@doi [\mnras] {10.1093/mnras/stac1918}, \href
  {https://ui.adsabs.harvard.edu/abs/2022MNRAS.515.3184H} {515, 3184}

\bibitem[\protect\citeauthoryear{{Huber}, {Stello}, {Bedding}, {Chaplin},
  {Arentoft}, {Quirion}  \& {Kjeldsen}}{{Huber}
  et~al.}{2009}]{Huber2009_sydpipeline}
{Huber} D.,  {Stello} D.,  {Bedding} T.~R.,  {Chaplin} W.~J.,  {Arentoft} T.,
  {Quirion} P.~O.,   {Kjeldsen} H.,  2009, \mn@doi [Communications in
  Asteroseismology] {10.48550/arXiv.0910.2764}, \href
  {https://ui.adsabs.harvard.edu/abs/2009CoAst.160...74H} {160, 74}

\bibitem[\protect\citeauthoryear{{Huber} et~al.,}{{Huber}
  et~al.}{2010}]{Huber2010_keplerrgs}
{Huber} D.,  et~al., 2010, \mn@doi [\apj] {10.1088/0004-637X/723/2/1607}, \href
  {https://ui.adsabs.harvard.edu/abs/2010ApJ...723.1607H} {723, 1607}

\bibitem[\protect\citeauthoryear{{Huber} et~al.,}{{Huber}
  et~al.}{2011a}]{Huber2011_solarvalues}
{Huber} D.,  et~al., 2011a, \mn@doi [\apj] {10.1088/0004-637X/731/2/94}, \href
  {https://ui.adsabs.harvard.edu/abs/2011ApJ...731...94H} {731, 94}

\bibitem[\protect\citeauthoryear{{Huber} et~al.,}{{Huber}
  et~al.}{2011b}]{Huber2011_testingscalingCHARA}
{Huber} D.,  et~al., 2011b, \mn@doi [\apj] {10.1088/0004-637X/743/2/143}, \href
  {https://ui.adsabs.harvard.edu/abs/2011ApJ...743..143H} {743, 143}

\bibitem[\protect\citeauthoryear{Hunter}{Hunter}{2007}]{matplotlib2007}
Hunter J.~D.,  2007, Computing in Science \& Engineering, 9, 90

\bibitem[\protect\citeauthoryear{{Iglesias} \& {Rogers}}{{Iglesias} \&
  {Rogers}}{1993}]{Iglesias1993}
{Iglesias} C.~A.,  {Rogers} F.~J.,  1993, \mn@doi [\apj] {10.1086/172958},
  \href {https://ui.adsabs.harvard.edu/abs/1993ApJ...412..752I} {412, 752}

\bibitem[\protect\citeauthoryear{{Iglesias} \& {Rogers}}{{Iglesias} \&
  {Rogers}}{1996}]{Iglesias1996}
{Iglesias} C.~A.,  {Rogers} F.~J.,  1996, \mn@doi [\apj] {10.1086/177381},
  \href {https://ui.adsabs.harvard.edu/abs/1996ApJ...464..943I} {464, 943}

\bibitem[\protect\citeauthoryear{{Irwin}}{{Irwin}}{2004}]{Irwin2004}
{Irwin} A.~W.,  2004, The FreeEOS Code for Calculating the Equation of State
  for Stellar Interiors, \url {http://freeeos.sourceforge.net/}

\bibitem[\protect\citeauthoryear{{Itoh}, {Hayashi}, {Nishikawa}  \&
  {Kohyama}}{{Itoh} et~al.}{1996}]{Itoh1996}
{Itoh} N.,  {Hayashi} H.,  {Nishikawa} A.,   {Kohyama} Y.,  1996, \mn@doi
  [\apjs] {10.1086/192264}, \href
  {https://ui.adsabs.harvard.edu/abs/1996ApJS..102..411I} {102, 411}

\bibitem[\protect\citeauthoryear{{Jackiewicz}}{{Jackiewicz}}{2021}]{Jackiewicz2021_review}
{Jackiewicz} J.,  2021, \mn@doi [Frontiers in Astronomy and Space Sciences]
  {10.3389/fspas.2020.595017}, \href
  {https://ui.adsabs.harvard.edu/abs/2021FrASS...7..102J} {7, 102}

\bibitem[\protect\citeauthoryear{{Jenkins} et~al.,}{{Jenkins}
  et~al.}{2010}]{Jenkins2010_keplernoise}
{Jenkins} J.~M.,  et~al., 2010, \mn@doi [\apjl] {10.1088/2041-8205/713/2/L120},
  \href {https://ui.adsabs.harvard.edu/abs/2010ApJ...713L.120J} {713, L120}

\bibitem[\protect\citeauthoryear{{Jermyn}, {Schwab}, {Bauer}, {Timmes}  \&
  {Potekhin}}{{Jermyn} et~al.}{2021}]{Jermyn2021}
{Jermyn} A.~S.,  {Schwab} J.,  {Bauer} E.,  {Timmes} F.~X.,   {Potekhin} A.~Y.,
   2021, \mn@doi [\apj] {10.3847/1538-4357/abf48e}, \href
  {https://ui.adsabs.harvard.edu/abs/2021ApJ...913...72J} {913, 72}

\bibitem[\protect\citeauthoryear{{Jermyn} et~al.,}{{Jermyn}
  et~al.}{2023}]{Jermyn2023}
{Jermyn} A.~S.,  et~al., 2023, \mn@doi [\apjs] {10.3847/1538-4365/acae8d},
  \href {https://ui.adsabs.harvard.edu/abs/2023ApJS..265...15J} {265, 15}

\bibitem[\protect\citeauthoryear{{J{\"o}nsson} et~al.,}{{J{\"o}nsson}
  et~al.}{2020}]{APOGEE_DR16}
{J{\"o}nsson} H.,  et~al., 2020, \mn@doi [\aj] {10.3847/1538-3881/aba592},
  \href {https://ui.adsabs.harvard.edu/abs/2020AJ....160..120J} {160, 120}

\bibitem[\protect\citeauthoryear{{Jorissen}, {Van Winckel}, {Siess}, {Escorza},
  {Pourbaix}  \& {Van Eck}}{{Jorissen} et~al.}{2020}]{Jorissen2020_SBpaper}
{Jorissen} A.,  {Van Winckel} H.,  {Siess} L.,  {Escorza} A.,  {Pourbaix} D.,
  {Van Eck} S.,  2020, \mn@doi [\aap] {10.1051/0004-6361/202037585}, \href
  {https://ui.adsabs.harvard.edu/abs/2020A&A...639A...7J} {639, A7}

\bibitem[\protect\citeauthoryear{{Kallinger}}{{Kallinger}}{2019}]{Kallinger_peakbaggingrepo}
{Kallinger} T.,  2019, \mn@doi [arXiv e-prints] {10.48550/arXiv.1906.09428},
  \href {https://ui.adsabs.harvard.edu/abs/2019arXiv190609428K} {p.
  arXiv:1906.09428}

\bibitem[\protect\citeauthoryear{{Kallinger} et~al.,}{{Kallinger}
  et~al.}{2010}]{Kallinger2010_corot}
{Kallinger} T.,  et~al., 2010, \mn@doi [\aap] {10.1051/0004-6361/200811437},
  \href {https://ui.adsabs.harvard.edu/abs/2010A&A...509A..77K} {509, A77}

\bibitem[\protect\citeauthoryear{{Kallinger} et~al.,}{{Kallinger}
  et~al.}{2012}]{Kallinger2012_epsilon}
{Kallinger} T.,  et~al., 2012, \mn@doi [\aap] {10.1051/0004-6361/201218854},
  \href {https://ui.adsabs.harvard.edu/abs/2012A&A...541A..51K} {541, A51}

\bibitem[\protect\citeauthoryear{{Kallinger} et~al.,}{{Kallinger}
  et~al.}{2014}]{Kallinger2014_granulation}
{Kallinger} T.,  et~al., 2014, \mn@doi [\aap] {10.1051/0004-6361/201424313},
  \href {https://ui.adsabs.harvard.edu/abs/2014A&A...570A..41K} {570, A41}

\bibitem[\protect\citeauthoryear{{Kim} \& {Chang}}{{Kim} \&
  {Chang}}{2021}]{Kim2021_width}
{Kim} K.-B.,  {Chang} H.-Y.,  2021, \mn@doi [\na]
  {10.1016/j.newast.2020.101522}, \href
  {https://ui.adsabs.harvard.edu/abs/2021NewA...8401522K} {84, 101522}

\bibitem[\protect\citeauthoryear{{Kinemuchi}, {Barclay}, {Fanelli}, {Pepper},
  {Still}  \& {Howell}}{{Kinemuchi}
  et~al.}{2012}]{Kinemuchi2012_demystifyingkepler}
{Kinemuchi} K.,  {Barclay} T.,  {Fanelli} M.,  {Pepper} J.,  {Still} M.,
  {Howell} S.~B.,  2012, \mn@doi [\pasp] {10.1086/667603}, \href
  {https://ui.adsabs.harvard.edu/abs/2012PASP..124..963K} {124, 963}

\bibitem[\protect\citeauthoryear{{Kjeldsen} \& {Bedding}}{{Kjeldsen} \&
  {Bedding}}{1995}]{Kjeldsen1995_scalingrelations}
{Kjeldsen} H.,  {Bedding} T.~R.,  1995, \mn@doi [\aap]
  {10.48550/arXiv.astro-ph/9403015}, \href
  {https://ui.adsabs.harvard.edu/abs/1995A&A...293...87K} {293, 87}

\bibitem[\protect\citeauthoryear{{Kjeldsen} \& {Bedding}}{{Kjeldsen} \&
  {Bedding}}{2011}]{Kjeldsen2011_amplitudescaling}
{Kjeldsen} H.,  {Bedding} T.~R.,  2011, \mn@doi [\aap]
  {10.1051/0004-6361/201116789}, \href
  {https://ui.adsabs.harvard.edu/abs/2011A&A...529L...8K} {529, L8}

\bibitem[\protect\citeauthoryear{{Kjeldsen} et~al.,}{{Kjeldsen}
  et~al.}{2005}]{Kjeldsen2005_alphacenB}
{Kjeldsen} H.,  et~al., 2005, \mn@doi [\apj] {10.1086/497530}, \href
  {https://ui.adsabs.harvard.edu/abs/2005ApJ...635.1281K} {635, 1281}

\bibitem[\protect\citeauthoryear{{Kjeldsen} et~al.,}{{Kjeldsen}
  et~al.}{2008}]{Kjeldsen2008_amplitudes}
{Kjeldsen} H.,  et~al., 2008, \mn@doi [\apj] {10.1086/589142}, \href
  {https://ui.adsabs.harvard.edu/abs/2008ApJ...682.1370K} {682, 1370}

\bibitem[\protect\citeauthoryear{{Kurtz}}{{Kurtz}}{2022}]{Kurtz2022_review}
{Kurtz} D.~W.,  2022, \mn@doi [\araa] {10.1146/annurev-astro-052920-094232},
  \href {https://ui.adsabs.harvard.edu/abs/2022ARA&A..60...31K} {60, 31}

\bibitem[\protect\citeauthoryear{{Langanke} \&
  {Mart{\'{\i}}nez-Pinedo}}{{Langanke} \&
  {Mart{\'{\i}}nez-Pinedo}}{2000}]{Langanke2000}
{Langanke} K.,  {Mart{\'{\i}}nez-Pinedo} G.,  2000, \mn@doi [Nuclear Physics A]
  {10.1016/S0375-9474(00)00131-7}, \href
  {https://ui.adsabs.harvard.edu/abs/2000NuPhA.673..481L} {673, 481}

\bibitem[\protect\citeauthoryear{{Lee}, {Offner}, {Hennebelle}, {Andr{\'e}},
  {Zinnecker}, {Ballesteros-Paredes}, {Inutsuka}  \& {Kruijssen}}{{Lee}
  et~al.}{2020}]{Lee2020_binaryfraction}
{Lee} Y.-N.,  {Offner} S. S.~R.,  {Hennebelle} P.,  {Andr{\'e}} P.,
  {Zinnecker} H.,  {Ballesteros-Paredes} J.,  {Inutsuka} S.-i.,   {Kruijssen}
  J.~M.~D.,  2020, \mn@doi [\ssr] {10.1007/s11214-020-00699-2}, \href
  {https://ui.adsabs.harvard.edu/abs/2020SSRv..216...70L} {216, 70}

\bibitem[\protect\citeauthoryear{{Li} et~al.,}{{Li}
  et~al.}{2023}]{Li2023_surfacecorrection}
{Li} Y.,  et~al., 2023, \mn@doi [\mnras] {10.1093/mnras/stad1445}, \href
  {https://ui.adsabs.harvard.edu/abs/2023MNRAS.523..916L} {523, 916}

\bibitem[\protect\citeauthoryear{{Lightkurve Collaboration}
  et~al.,}{{Lightkurve Collaboration} et~al.}{2018a}]{lightkurve}
{Lightkurve Collaboration} et~al., 2018a, {Lightkurve: Kepler and TESS time
  series analysis in Python}, Astrophysics Source Code Library (\mn@eprint
  {ascl} {1812.013})

\bibitem[\protect\citeauthoryear{{Lightkurve Collaboration}
  et~al.,}{{Lightkurve Collaboration} et~al.}{2018b}]{lightkurve2018}
{Lightkurve Collaboration} et~al., 2018b, {Lightkurve: Kepler and TESS time
  series analysis in Python}, Astrophysics Source Code Library (\mn@eprint
  {ascl} {1812.013})

\bibitem[\protect\citeauthoryear{{Loi} \& {Papaloizou}}{{Loi} \&
  {Papaloizou}}{2018}]{Loi2018_magneticsuppression_theory}
{Loi} S.~T.,  {Papaloizou} J. C.~B.,  2018, \mn@doi [\mnras]
  {10.1093/mnras/sty917}, \href
  {https://ui.adsabs.harvard.edu/abs/2018MNRAS.477.5338L} {477, 5338}

\bibitem[\protect\citeauthoryear{{Lomb}}{{Lomb}}{1976}]{Lomb1976_lombscargle}
{Lomb} N.~R.,  1976, \mn@doi [\apss] {10.1007/BF00648343}, \href
  {https://ui.adsabs.harvard.edu/abs/1976Ap&SS..39..447L} {39, 447}

\bibitem[\protect\citeauthoryear{{Mackereth} et~al.,}{{Mackereth}
  et~al.}{2021}]{Mackereth2021_tess_cvz}
{Mackereth} J.~T.,  et~al., 2021, \mn@doi [\mnras] {10.1093/mnras/stab098},
  \href {https://ui.adsabs.harvard.edu/abs/2021MNRAS.502.1947M} {502, 1947}

\bibitem[\protect\citeauthoryear{{Mathur} et~al.,}{{Mathur}
  et~al.}{2011}]{Mathur2011_granulation}
{Mathur} S.,  et~al., 2011, \mn@doi [\apj] {10.1088/0004-637X/741/2/119}, \href
  {https://ui.adsabs.harvard.edu/abs/2011ApJ...741..119M} {741, 119}

\bibitem[\protect\citeauthoryear{{Mathur} et~al.,}{{Mathur}
  et~al.}{2017}]{Mathur2017_temperatures}
{Mathur} S.,  et~al., 2017, \mn@doi [\apjs] {10.3847/1538-4365/229/2/30}, \href
  {https://ui.adsabs.harvard.edu/abs/2017ApJS..229...30M} {229, 30}

\bibitem[\protect\citeauthoryear{{Miglio} et~al.,}{{Miglio}
  et~al.}{2012}]{Miglio2012_clustermassloss}
{Miglio} A.,  et~al., 2012, \mn@doi [\mnras]
  {10.1111/j.1365-2966.2011.19859.x}, \href
  {https://ui.adsabs.harvard.edu/abs/2012MNRAS.419.2077M} {419, 2077}

\bibitem[\protect\citeauthoryear{{Moe} \& {Di Stefano}}{{Moe} \& {Di
  Stefano}}{2017}]{Moe2017_multiplicity}
{Moe} M.,  {Di Stefano} R.,  2017, \mn@doi [\apjs] {10.3847/1538-4365/aa6fb6},
  \href {https://ui.adsabs.harvard.edu/abs/2017ApJS..230...15M} {230, 15}

\bibitem[\protect\citeauthoryear{{Molenda-{\.Z}akowicz}, {Brogaard},
  {Niemczura}, {Bergemann}, {Frasca}, {Arentoft}  \&
  {Grundahl}}{{Molenda-{\.Z}akowicz}
  et~al.}{2014}]{MolendaZakowicz2014_NGC6811SB}
{Molenda-{\.Z}akowicz} J.,  {Brogaard} K.,  {Niemczura} E.,  {Bergemann} M.,
  {Frasca} A.,  {Arentoft} T.,   {Grundahl} F.,  2014, \mn@doi [\mnras]
  {10.1093/mnras/stu1934}, \href
  {https://ui.adsabs.harvard.edu/abs/2014MNRAS.445.2446M} {445, 2446}

\bibitem[\protect\citeauthoryear{{Montalb{\'a}n}, {Miglio}, {Noels}, {Dupret},
  {Scuflaire}  \& {Ventura}}{{Montalb{\'a}n}
  et~al.}{2013}]{Montalban2013_pspacingconvection}
{Montalb{\'a}n} J.,  {Miglio} A.,  {Noels} A.,  {Dupret} M.~A.,  {Scuflaire}
  R.,   {Ventura} P.,  2013, \mn@doi [\apj] {10.1088/0004-637X/766/2/118},
  \href {https://ui.adsabs.harvard.edu/abs/2013ApJ...766..118M} {766, 118}

\bibitem[\protect\citeauthoryear{{Mosser} et~al.,}{{Mosser}
  et~al.}{2011a}]{Mosser2011_universalpattern}
{Mosser} B.,  et~al., 2011a, \mn@doi [\aap] {10.1051/0004-6361/201015440},
  \href {https://ui.adsabs.harvard.edu/abs/2011A&A...525L...9M} {525, L9}

\bibitem[\protect\citeauthoryear{{Mosser} et~al.,}{{Mosser}
  et~al.}{2011b}]{Mosser2011_corot_mixed_modes}
{Mosser} B.,  et~al., 2011b, \mn@doi [\aap] {10.1051/0004-6361/201116825},
  \href {https://ui.adsabs.harvard.edu/abs/2011A&A...532A..86M} {532, A86}

\bibitem[\protect\citeauthoryear{{Mosser} et~al.,}{{Mosser}
  et~al.}{2012}]{Mosser2012_powerexcess}
{Mosser} B.,  et~al., 2012, \mn@doi [\aap]
  {10.1051/0004-6361/20111735210.1086/141952}, \href
  {https://ui.adsabs.harvard.edu/abs/2012A&A...537A..30M} {537, A30}

\bibitem[\protect\citeauthoryear{{Mosser} et~al.,}{{Mosser}
  et~al.}{2017}]{Mosser2017_suppression_mixedmodes}
{Mosser} B.,  et~al., 2017, \mn@doi [\aap] {10.1051/0004-6361/201629494}, \href
  {https://ui.adsabs.harvard.edu/abs/2017A&A...598A..62M} {598, A62}

\bibitem[\protect\citeauthoryear{{Murphy}}{{Murphy}}{2012}]{Murphy2012_Keplercharacteristics}
{Murphy} S.~J.,  2012, \mn@doi [\mnras] {10.1111/j.1365-2966.2012.20644.x},
  \href {https://ui.adsabs.harvard.edu/abs/2012MNRAS.422..665M} {422, 665}

\bibitem[\protect\citeauthoryear{{Oda}, {Hino}, {Muto}, {Takahara}  \&
  {Sato}}{{Oda} et~al.}{1994}]{Oda1994}
{Oda} T.,  {Hino} M.,  {Muto} K.,  {Takahara} M.,   {Sato} K.,  1994, \mn@doi
  [Atomic Data and Nuclear Data Tables] {10.1006/adnd.1994.1007}, \href
  {https://ui.adsabs.harvard.edu/abs/1994ADNDT..56..231O} {56, 231}

\bibitem[\protect\citeauthoryear{{Ong} \& {Gehan}}{{Ong} \&
  {Gehan}}{2023}]{Ong2023_pspacingalgo}
{Ong} J.~M.~J.,  {Gehan} C.,  2023, \mn@doi [\apj] {10.3847/1538-4357/acbf2f},
  \href {https://ui.adsabs.harvard.edu/abs/2023ApJ...946...92O} {946, 92}

\bibitem[\protect\citeauthoryear{{Paxton}, {Bildsten}, {Dotter}, {Herwig},
  {Lesaffre}  \& {Timmes}}{{Paxton} et~al.}{2011}]{Paxton2011}
{Paxton} B.,  {Bildsten} L.,  {Dotter} A.,  {Herwig} F.,  {Lesaffre} P.,
  {Timmes} F.,  2011, \mn@doi [\apjs] {10.1088/0067-0049/192/1/3}, \href
  {https://ui.adsabs.harvard.edu/abs/2011ApJS..192....3P} {192, 3}

\bibitem[\protect\citeauthoryear{{Paxton} et~al.,}{{Paxton}
  et~al.}{2013}]{Paxton2013}
{Paxton} B.,  et~al., 2013, \mn@doi [\apjs] {10.1088/0067-0049/208/1/4}, \href
  {https://ui.adsabs.harvard.edu/abs/2013ApJS..208....4P} {208, 4}

\bibitem[\protect\citeauthoryear{{Paxton} et~al.,}{{Paxton}
  et~al.}{2015}]{Paxton2015}
{Paxton} B.,  et~al., 2015, \mn@doi [\apjs] {10.1088/0067-0049/220/1/15}, \href
  {https://ui.adsabs.harvard.edu/abs/2015ApJS..220...15P} {220, 15}

\bibitem[\protect\citeauthoryear{{Paxton} et~al.,}{{Paxton}
  et~al.}{2018}]{Paxton2018}
{Paxton} B.,  et~al., 2018, \mn@doi [\apjs] {10.3847/1538-4365/aaa5a8}, \href
  {https://ui.adsabs.harvard.edu/abs/2018ApJS..234...34P} {234, 34}

\bibitem[\protect\citeauthoryear{{Paxton} et~al.,}{{Paxton}
  et~al.}{2019}]{Paxton2019}
{Paxton} B.,  et~al., 2019, \mn@doi [\apjs] {10.3847/1538-4365/ab2241}, \href
  {https://ui.adsabs.harvard.edu/abs/2019ApJS..243...10P} {243, 10}

\bibitem[\protect\citeauthoryear{{Pinsonneault} et~al.,}{{Pinsonneault}
  et~al.}{2018}]{Pinsonneault2018_apokasc2}
{Pinsonneault} M.~H.,  et~al., 2018, \mn@doi [\apjs]
  {10.3847/1538-4365/aaebfd}, \href
  {https://ui.adsabs.harvard.edu/abs/2018ApJS..239...32P} {239, 32}

\bibitem[\protect\citeauthoryear{{Potekhin} \& {Chabrier}}{{Potekhin} \&
  {Chabrier}}{2010}]{Potekhin2010}
{Potekhin} A.~Y.,  {Chabrier} G.,  2010, \mn@doi [Contributions to Plasma
  Physics] {10.1002/ctpp.201010017}, \href
  {https://ui.adsabs.harvard.edu/abs/2010CoPP...50...82P} {50, 82}

\bibitem[\protect\citeauthoryear{{Poutanen}}{{Poutanen}}{2017}]{Poutanen2017}
{Poutanen} J.,  2017, \mn@doi [\apj] {10.3847/1538-4357/835/2/119}, \href
  {https://ui.adsabs.harvard.edu/abs/2017ApJ...835..119P} {835, 119}

\bibitem[\protect\citeauthoryear{{Pr{\v{s}}a} et~al.,}{{Pr{\v{s}}a}
  et~al.}{2016}]{Prsa2016_solar_iau}
{Pr{\v{s}}a} A.,  et~al., 2016, \mn@doi [\aj] {10.3847/0004-6256/152/2/41},
  \href {https://ui.adsabs.harvard.edu/abs/2016AJ....152...41P} {152, 41}

\bibitem[\protect\citeauthoryear{{Rogers} \& {Nayfonov}}{{Rogers} \&
  {Nayfonov}}{2002}]{Rogers2002}
{Rogers} F.~J.,  {Nayfonov} A.,  2002, \mn@doi [\apj] {10.1086/341894}, \href
  {https://ui.adsabs.harvard.edu/abs/2002ApJ...576.1064R} {576, 1064}

\bibitem[\protect\citeauthoryear{{Saumon}, {Chabrier}  \& {van Horn}}{{Saumon}
  et~al.}{1995}]{Saumon1995}
{Saumon} D.,  {Chabrier} G.,   {van Horn} H.~M.,  1995, \mn@doi [\apjs]
  {10.1086/192204}, \href
  {https://ui.adsabs.harvard.edu/abs/1995ApJS...99..713S} {99, 713}

\bibitem[\protect\citeauthoryear{{Scargle}}{{Scargle}}{1982}]{Scargle1982_lombscargle}
{Scargle} J.~D.,  1982, \mn@doi [\apj] {10.1086/160554}, \href
  {https://ui.adsabs.harvard.edu/abs/1982ApJ...263..835S} {263, 835}

\bibitem[\protect\citeauthoryear{{Schonhut-Stasik} et~al.,}{{Schonhut-Stasik}
  et~al.}{2020}]{Schonhut-Stasik2020_binarysuppression}
{Schonhut-Stasik} J.,  et~al., 2020, \mn@doi [\apj] {10.3847/1538-4357/ab50c3},
  \href {https://ui.adsabs.harvard.edu/abs/2020ApJ...888...34S} {888, 34}

\bibitem[\protect\citeauthoryear{{Sharma}, {Stello}, {Bland-Hawthorn}, {Huber}
  \& {Bedding}}{{Sharma} et~al.}{2016}]{Sharma2016_fdnu}
{Sharma} S.,  {Stello} D.,  {Bland-Hawthorn} J.,  {Huber} D.,   {Bedding}
  T.~R.,  2016, \mn@doi [\apj] {10.3847/0004-637X/822/1/15}, \href
  {https://ui.adsabs.harvard.edu/abs/2016ApJ...822...15S} {822, 15}

\bibitem[\protect\citeauthoryear{{Stello} \& {Sharma}}{{Stello} \&
  {Sharma}}{2022}]{Stello2022_asfgridextension}
{Stello} D.,  {Sharma} S.,  2022, \mn@doi [Research Notes of the American
  Astronomical Society] {10.3847/2515-5172/ac8b12}, \href
  {https://ui.adsabs.harvard.edu/abs/2022RNAAS...6..168S} {6, 168}

\bibitem[\protect\citeauthoryear{{Stello}, {Bruntt}, {Preston}  \&
  {Buzasi}}{{Stello} et~al.}{2008}]{Stello2008_wire}
{Stello} D.,  {Bruntt} H.,  {Preston} H.,   {Buzasi} D.,  2008, \mn@doi [\apjl]
  {10.1086/528936}, \href
  {https://ui.adsabs.harvard.edu/abs/2008ApJ...674L..53S} {674, L53}

\bibitem[\protect\citeauthoryear{{Stello} et~al.,}{{Stello}
  et~al.}{2011}]{Stello2011_keplerclusters}
{Stello} D.,  et~al., 2011, \mn@doi [\apj] {10.1088/0004-637X/739/1/13}, \href
  {https://ui.adsabs.harvard.edu/abs/2011ApJ...739...13S} {739, 13}

\bibitem[\protect\citeauthoryear{{Stello} et~al.,}{{Stello}
  et~al.}{2013}]{Stello2013_pspacingtracks}
{Stello} D.,  et~al., 2013, \mn@doi [\apjl] {10.1088/2041-8205/765/2/L41},
  \href {https://ui.adsabs.harvard.edu/abs/2013ApJ...765L..41S} {765, L41}

\bibitem[\protect\citeauthoryear{{Stello}, {Cantiello}, {Fuller}, {Garcia}  \&
  {Huber}}{{Stello} et~al.}{2016a}]{Stello2016_visPASA}
{Stello} D.,  {Cantiello} M.,  {Fuller} J.,  {Garcia} R.~A.,   {Huber} D.,
  2016a, \mn@doi [\pasa] {10.1017/pasa.2016.9}, \href
  {https://ui.adsabs.harvard.edu/abs/2016PASA...33...11S} {33, e011}

\bibitem[\protect\citeauthoryear{{Stello}, {Cantiello}, {Fuller}, {Huber},
  {Garc{\'\i}a}, {Bedding}, {Bildsten}  \& {Silva Aguirre}}{{Stello}
  et~al.}{2016b}]{Stello2016_visNature}
{Stello} D.,  {Cantiello} M.,  {Fuller} J.,  {Huber} D.,  {Garc{\'\i}a} R.~A.,
  {Bedding} T.~R.,  {Bildsten} L.,   {Silva Aguirre} V.,  2016b, \mn@doi [\nat]
  {10.1038/nature16171}, \href
  {https://ui.adsabs.harvard.edu/abs/2016Natur.529..364S} {529, 364}

\bibitem[\protect\citeauthoryear{{Stello} et~al.,}{{Stello}
  et~al.}{2016c}]{Stello2016_masslossM67}
{Stello} D.,  et~al., 2016c, \mn@doi [\apj] {10.3847/0004-637X/832/2/133},
  \href {https://ui.adsabs.harvard.edu/abs/2016ApJ...832..133S} {832, 133}

\bibitem[\protect\citeauthoryear{{Stumpe} et~al.,}{{Stumpe}
  et~al.}{2012}]{Stumpe2012_PDCKepler}
{Stumpe} M.~C.,  et~al., 2012, \mn@doi [\pasp] {10.1086/667698}, \href
  {https://ui.adsabs.harvard.edu/abs/2012PASP..124..985S} {124, 985}

\bibitem[\protect\citeauthoryear{{Timmes} \& {Swesty}}{{Timmes} \&
  {Swesty}}{2000}]{Timmes2000}
{Timmes} F.~X.,  {Swesty} F.~D.,  2000, \mn@doi [\apjs] {10.1086/313304}, \href
  {https://ui.adsabs.harvard.edu/abs/2000ApJS..126..501T} {126, 501}

\bibitem[\protect\citeauthoryear{{Ulrich}}{{Ulrich}}{1986}]{Ulrich1986_dnuscaling}
{Ulrich} R.~K.,  1986, \mn@doi [\apjl] {10.1086/184700}, \href
  {https://ui.adsabs.harvard.edu/abs/1986ApJ...306L..37U} {306, L37}

\bibitem[\protect\citeauthoryear{Virtanen et~al.,}{Virtanen
  et~al.}{2020}]{scipy2020}
Virtanen P.,  et~al., 2020, \mn@doi [Nature Methods]
  {10.1038/s41592-019-0686-2}, \href {https://rdcu.be/b08Wh} {17, 261}

\bibitem[\protect\citeauthoryear{{Vrard} et~al.,}{{Vrard}
  et~al.}{2015}]{Vrard2015_glitches}
{Vrard} M.,  et~al., 2015, \mn@doi [\aap] {10.1051/0004-6361/201425064}, \href
  {https://ui.adsabs.harvard.edu/abs/2015A&A...579A..84V} {579, A84}

\bibitem[\protect\citeauthoryear{{Vrard}, {Mosser}  \& {Samadi}}{{Vrard}
  et~al.}{2016}]{Vrard2016_pspacings}
{Vrard} M.,  {Mosser} B.,   {Samadi} R.,  2016, \mn@doi [\aap]
  {10.1051/0004-6361/201527259}, \href
  {https://ui.adsabs.harvard.edu/abs/2016A&A...588A..87V} {588, A87}

\bibitem[\protect\citeauthoryear{{White}, {Bedding}, {Stello},
  {Christensen-Dalsgaard}, {Huber}  \& {Kjeldsen}}{{White}
  et~al.}{2011}]{White2011_solarlike_models}
{White} T.~R.,  {Bedding} T.~R.,  {Stello} D.,  {Christensen-Dalsgaard} J.,
  {Huber} D.,   {Kjeldsen} H.,  2011, \mn@doi [\apj]
  {10.1088/0004-637X/743/2/161}, \href
  {https://ui.adsabs.harvard.edu/abs/2011ApJ...743..161W} {743, 161}

\bibitem[\protect\citeauthoryear{{Yu}, {Huber}, {Bedding}, {Stello}, {Hon},
  {Murphy}  \& {Khanna}}{{Yu} et~al.}{2018}]{Yu2018_keplercatalog}
{Yu} J.,  {Huber} D.,  {Bedding} T.~R.,  {Stello} D.,  {Hon} M.,  {Murphy}
  S.~J.,   {Khanna} S.,  2018, \mn@doi [\apjs] {10.3847/1538-4365/aaaf74},
  \href {https://ui.adsabs.harvard.edu/abs/2018ApJS..236...42Y} {236, 42}

\bibitem[\protect\citeauthoryear{{Yu}, {Hekker}, {Bedding}, {Stello}, {Huber},
  {Gizon}, {Khanna}  \& {Bi}}{{Yu} et~al.}{2021}]{Yu2021_luminosrg_massloss}
{Yu} J.,  {Hekker} S.,  {Bedding} T.~R.,  {Stello} D.,  {Huber} D.,  {Gizon}
  L.,  {Khanna} S.,   {Bi} S.,  2021, \mn@doi [\mnras]
  {10.1093/mnras/staa3970}, \href
  {https://ui.adsabs.harvard.edu/abs/2021MNRAS.501.5135Y} {501, 5135}

\bibitem[\protect\citeauthoryear{{Zinn}, {Pinsonneault}, {Huber}, {Stello},
  {Stassun}  \& {Serenelli}}{{Zinn}
  et~al.}{2019}]{Zinn2019_testingscalingrelation_gaia}
{Zinn} J.~C.,  {Pinsonneault} M.~H.,  {Huber} D.,  {Stello} D.,  {Stassun} K.,
   {Serenelli} A.,  2019, \mn@doi [\apj] {10.3847/1538-4357/ab44a9}, \href
  {https://ui.adsabs.harvard.edu/abs/2019ApJ...885..166Z} {885, 166}

\makeatother
\end{thebibliography}

\bsp	
\label{lastpage}
\end{document}